\documentclass[superscriptaddress,twocolumn,10pt,pra,aps,showpacs,longbibliography]{revtex4-1}
\usepackage{float}
\floatstyle{boxed}
\usepackage{wrapfig}
\usepackage{array}
\usepackage{graphicx}
\usepackage{amsbsy}
\usepackage[utf8x]{inputenc}
\usepackage{epstopdf}
\usepackage{amsmath,amssymb,amsfonts}

\def \am{\hat a_1}
\def \ap{\hat a_1^{\dagger}}

\newcommand{\PT}{${\cal PT}$}

\newcommand{\ket}[1]{|#1\rangle}

\renewcommand{\eqref}[1]{\mbox{Eq.~(\ref{#1})}}

\newcommand{\be}{\begin{equation}}
\newcommand{\ee}{\end{equation}}
\newcommand{\bea}{\begin{eqnarray}}
\newcommand{\eea}{\end{eqnarray}}
\newcommand{\Hef}{\hat H_{\rm eff}}

\usepackage{url}
\usepackage[colorlinks]{hyperref}
\hypersetup{
    plainpages=true,
    breaklinks=true,
    hypertexnames=false,
    pageanchor=true,
    colorlinks=true,
    linkcolor={blue},
    citecolor={red},
    urlcolor={blue},
    anchorcolor={black}
}

\usepackage[normalem]{ulem} 

\begin{document}
\author{Ievgen I. Arkhipov}
\email{ievgen.arkhipov@upol.cz} \affiliation{Joint Laboratory of
Optics of Palack\'y University and Institute of Physics of CAS,
Faculty of Science, Palack\'y University, 17. listopadu 12, 771 46
Olomouc, Czech Republic}

\author{Adam Miranowicz}
\email{miran@amu.edu.pl} \affiliation{Faculty of Physics, Adam
Mickiewicz University, PL-61-614 Poznan, Poland}
\affiliation{Theoretical Quantum Physics Laboratory, RIKEN Cluster
for Pioneering Research, Wako-shi, Saitama 351-0198, Japan}
\author{Fabrizio Minganti}
\email{fabrizio.minganti@riken.jp} \affiliation{Theoretical
Quantum Physics Laboratory, RIKEN Cluster for Pioneering Research,
Wako-shi, Saitama 351-0198, Japan}
\author{Franco Nori}
\email{fnori@riken.jp} \affiliation{Theoretical Quantum Physics
Laboratory, RIKEN Cluster for Pioneering Research, Wako-shi,
Saitama 351-0198, Japan} \affiliation{Physics Department, The
University of Michigan, Ann Arbor, Michigan 48109-1040, USA}

\title{Quantum and semiclassical exceptional points of a linear system of coupled
cavities with losses and gain within the Scully-Lamb laser theory}

\begin{abstract}
In the past few decades, many works have been devoted to the study
of exceptional points (EPs), i.e., exotic degeneracies of
non-Hermitian systems. The usual approach in those studies
involves the introduction of a phenomenological effective
non-Hermitian Hamiltonian (NHH), where the gain and losses are
incorporated as the imaginary frequencies of fields and from
which the Hamiltonian EPs (HEPs) are derived. Although this
approach can provide valid equations of motion for the fields in
the classical limit, its application in the derivation of EPs in
the quantum regime is questionable. Recently, a framework
[Minganti {\it et al.},
\href{https://doi.org/10.1103/PhysRevA.100.062131}{Phys. Rev. A {\bf 100}, 062131 (2019)}], which
allows one to determine quantum EPs from a Liouvillian  EP (LEP), rather
than from  an NHH, has been proposed. Compared to the NHHs, a
Liouvillian naturally includes quantum noise effects via
quantum-jump terms, thus allowing one to consistently determine its
EPs purely in the quantum regime. In this work we study a
non-Hermitian system consisting of coupled cavities with
unbalanced gain and losses,  where the gain is far from
saturation, i.e, the system is assumed to be linear. We apply both
formalisms, based on an NHH and a Liouvillian within the
Scully-Lamb laser theory,  to determine and compare the
corresponding HEPs and LEPs in the semiclassical and quantum
regimes. {Our results indicate that, although the overall spectral
properties of the NHH and the corresponding Liouvillian for a
given system can differ substantially, their LEPs and HEPs occur
for the same combination of system parameters.}
\end{abstract}

\date{\today}

\maketitle

\section{Introduction}

Non-Hermiticity plays a crucial role in the study of the
dynamics of quantum systems. Non-Hermiticity refers to the systems
described by Hamiltonians that are non-Hermitian, i.e., the energy
spectra are represented by complex values. The positive or
negative imaginary parts of the eigenvalues of a non-Hermitian
Hamiltonian (NHH) indicate that a given  system undergoes either
amplification or dissipation processes, respectively. The best
known examples of non-Hermitian systems are open quantum systems,
where a quantum system of interest interacts with an environment,
where the latter induces decoherence of the former.

Recently, a new surge of interest in non-Hermitian systems has
been triggered by the discovery of a class of non-Hermitian
Hamiltonians, which commute with a parity-time ($\cal PT$)
operator, with real eigenvalues~\cite{Bender1998}.
Initially,  $\cal PT$-symmetric systems were merely an object of
mathematical interest, as there was  little understanding on how
to implement such systems in practice. It was only later realized
that $\cal PT$-symmetry can be carried out in photonics, due to
the analogy of the Schr\"{o}dinger equation in quantum mechanics
and the paraxial Maxwell equation in classical
physics~\cite{Ozdemir2019,Miri2019,Liang2017,El-Ganainy2018,ChristodoulidesBook}.
In the latter case, {this analogy can be explored} by making
the profile of the real and imaginary parts of the optical index
of a medium symmetric and asymmetric, respectively. {Thus, one
can obtain the system, which exhibits a $\cal PT$ symmetry-like
behavior, by properly balancing gain and losses of the system.}

One of the most peculiar properties of non-Hermitian systems, in
particular those which are  $\cal PT$-symmetric, is the presence
of the so-called exceptional points (EPs), i.e., system
degeneracies, where both eigenvalues and their corresponding
eigenvectors of an NHH coincide. The behavior of physical systems
near EPs can lead to the observation of nontrivial phenomena in
photonics~\cite{Ozdemir2019,Miri2019}. These include:
unidirectional invisibility~\cite{Lin2011,Regen2012},  lasers with
and enhanced-mode selectivity~\cite{Feng2014,Hodaei2014},
low-power nonreciprocal light transmission
\cite{Peng2014,Chang2014}, thresholdless phonon lasers
\cite{Jing2014,Lu2017}, enhanced light-matter interactions
\cite{Liu2016,Chen2017,Hoda2017},  and loss-induced lasing
\cite{Brands2014a,Peng2014a}. Exceptional points have been discussed in
electronics~\cite{Schindler2011}, optomechanics
\cite{Jing2014,Harris2016,Jing2017}, acoustics
\cite{Zhu2014,Alu2015}, plasmonics~\cite{Benisty2011}, and
metamaterials~\cite{Kang2013}.  The concept of EPs has been
successfully applied in the description of dynamical quantum phase
transitions and topological phases of matter in open quantum
systems (see,
e.g.,~\cite{LeykamPRL17,GonzalesPRB17,HuPRB17,GaoPRL18,LiuPRL19,Zhou2018,BliokhNat19,MoosSciPost19,Ge2019,Yoshida2019}).

So far, the concept of EPs in photonics has been mostly exploited
within the framework of effective NHHs, where gain and losses are
introduced phenomenologically into the Hamiltonians as the
imaginary part of the field frequencies. The use of such an
approach can be justified in the semiclassical regime, i.e., when
considering intense classical fields.  However, that approach
can fail in the quantum regime, where the explicit inclusion of
quantum noise and spontaneous emission becomes necessary. Needless
to say, quantum noise leads to symmetry breaking, in particular,
$\cal PT$-symmetry breaking~\cite{Scheel2018}. The quantum noise
in a system can be precisely simulated by either the  master
equation (ME)~\cite{ScullyLambBook,AgarwalBook} or the quantum
trajectory method~\cite{HarocheBook,Gea1998}. Of course, one can
also resort to quantum Langevin forces within the framework of an
NHH, but such an approach bears a phenomenological character and
in some cases, can lead to erroneous
results~\cite{ScullyLambBook,Zhou2019}.

The ME with a Liouvillian superoperator captures all the dynamics
of an open quantum system with {Markovian} gain and losses.
Recently, the concept of EPs based on the degeneracies  of the
Liouvillian rather than of an effective NHH was introduced in Refs.~\cite{Minganti2019,Prosen2012}. The study of the spectrum of a
Liouvillian provides a framework for the investigation of the
properties of non-Hermitian systems and their EPs in a rigorous
quantum
approach~\cite{Minganti2018,Macieszczak2016,Hatano2019,Albert2014,Sarandy2005,Prosen2010,Prosen2012}.

In this work, we focus on a {\it linear} non-Hermitian system
consisting of two coupled active and passive cavities with gain
and loss, respectively. The system is assumed to be linear,
because  the active cavity is assumed  to operate far below the
lasing threshold.

We study and compare EPs derived from two different formalisms
based on  an effective NHH and  a Liouvillian. Furthermore, we
analyze Hamiltonian EPs (HEPs) andLiouvillian EPs (LEPs) in both semiclassical, {(i.e., when
quantum jumps can be effectively ignored, which usually is the
case for systems with} large mean photon number, $\langle\hat
n\rangle\gg1$, and quantum regimes, {(i.e., when quantum jumps
cannot be ignored, e.g., for quantum systems with} very small mean
photon number $\langle\hat n\rangle\ll1$. In both regimes, we
treat the fields as $q$ numbers.

In the semiclassical regime, we
determine HEPs from the eigenspectra of the Hamiltonian, which is
written in a finite-matrix form, whereas LEPs are derived via a
two-time correlation function (TTCF), since a direct
diagonalization of the Liouvillian is almost impossible for
$\langle\hat n\rangle\gg1$.
In contrast, in the quantum single-photon limit, both
Hamiltonian and Liouvillian can be represented as finite matrices
thus  allowing us to determine their HEPs and LEPs solely from
their eigenspectra.

Our results indicate that the same combination
of system parameters leads to the occurrence of HEPs and LEPs in either regime.
Remarkably, the overall spectral properties of the
Liouvillian and NHH can differ substantially. {Indeed, we find that LEPs can be
of higher order than that of the corresponding HEPs.}

Additionally, when
considering the semiclassical regime, we provide a comparison of
LEPs determined from both TTCFs and spectral bifurcation points
(SBPs) of power spectra. Thus, we present a comparison of LEPs
defined in two complementary domains. This comparison reveals
that, in general, only  TTCFs can be used for identifying a true
LEP in the semiclassical limit.

The paper is organized as follows. In Sec.~II we introduce both
 Liouvillian and effective NHH for the linear system of
coupled active and passive cavities. In Secs.~III and IV we study
and compare HEPs and LEPs in the semiclassical and quantum
regimes, respectively. We summarize and draw conclusions in Sec.~V.

Through the text of this paper we deal with several abbreviations.
Therefore, in order to avoid any confusion when encountering them,
we list all of them in Table~I.
\begin{table}[t!]
\begin{center}
\begin{tabular}{ c|c}
Full name & Abbreviation \\
\hline \hline
Non-Hermitian Hamiltonian & NHH \\
\hline
Exceptional point  &  EP \\
\hline
Hamiltonian exceptional point & HEP \\
(an EP of an NHH) &  \\ \hline
Liouvillian exceptional point & LEP \\
(an EP of a Liouvillian) &  \\ \hline
Spectral bifurcation point & SBP \\
(a bifurcation point of a power spectrum) & \\ \hline
Master equation & ME \\
\hline
Two-time correlation function & TTCF \\
\end{tabular}
\caption{Abbreviations used in this paper.}
\end{center}
\end{table}

\section{General Theory of the Scully-Lamb model in the quantum limit}

\begin{figure}[tb] 
    \includegraphics[width=0.4\textwidth]{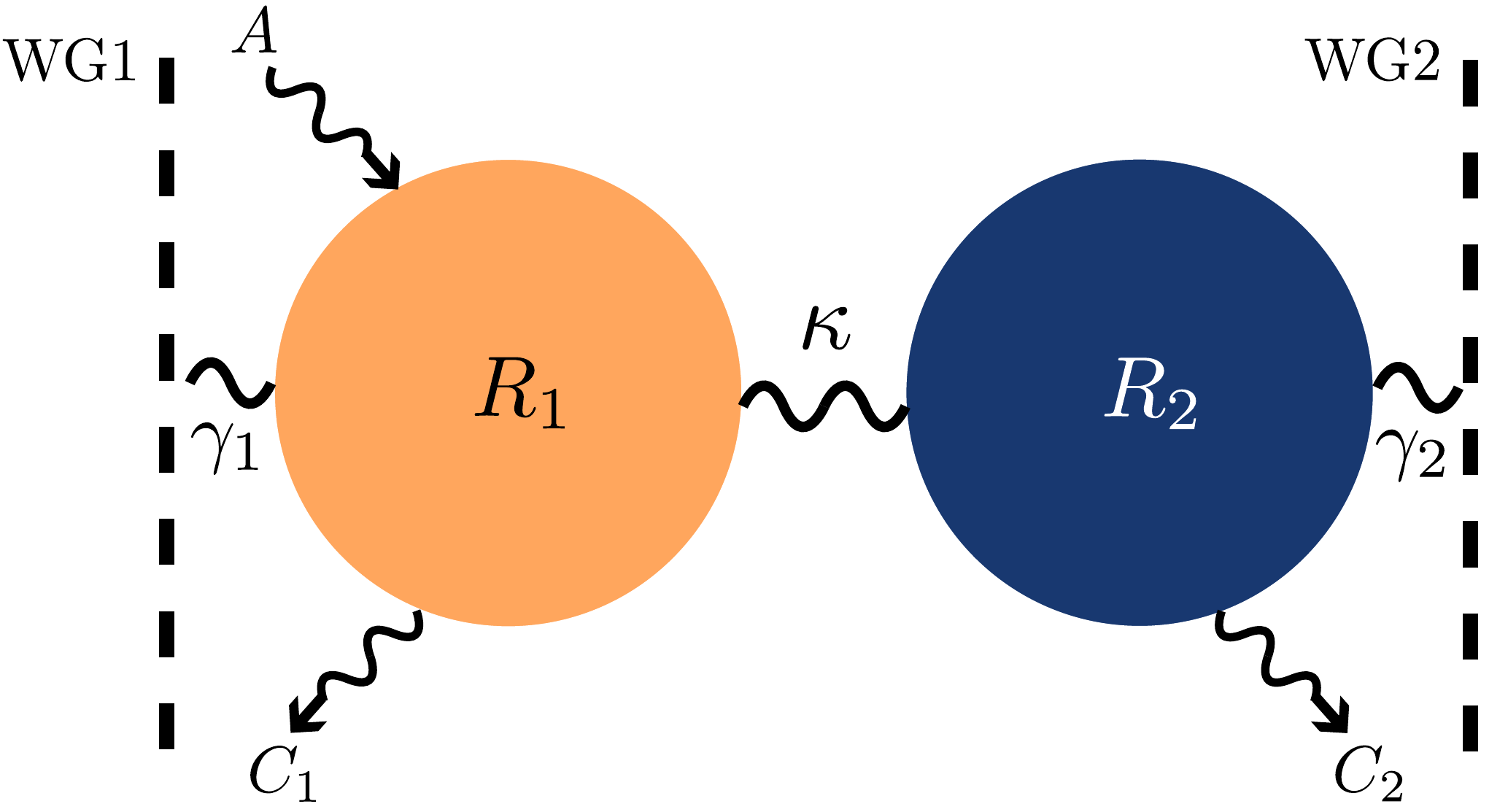}
\caption{{Setup of the system of linearly coupled active and
passive resonators. The active cavity $R_1$ has a gain rate $A$
and the total loss rate $\Gamma_1=C_1+\gamma_1$, consisting of the
intrinsic loss rate $C_1$ and the loss rate $\gamma_1$ due to the
coupling of $R_1$ to the waveguide WG1. The passive cavity $R_2$
has a total leakage rate $\Gamma_2=C_2+\gamma_2$, with $C_2$ and
$\gamma_2$ being an intrinsic loss and a leakage loss to the
waveguide WG2, respectively. The coupling strength between the
active $R_1$ and passive $R_2$ resonators   is denoted by
$\kappa$.}}\label{fig1}
\end{figure}

The object of our study is the system of two coupled cavities,
sketched in Fig.~\ref{fig1}, where one cavity is active, i.e., it
can provide gain for fields, and the other cavity  is passive, i.e.,
it induces only losses. Additionally, each resonator is coupled to
a waveguide  (see Fig.~\ref{fig1}).

The Hamiltonian of the system  can be written as
\begin{eqnarray}\label{H}  
\hat H = &&\sum\limits_{k=1}^2\hbar\omega_k\hat a_k^{\dagger}\hat
a_k+i\hbar\kappa(\hat a_1\hat a_2^{\dagger} -{\rm H.c.}),
\end{eqnarray}
where {$\hat a_k$ ($\hat a_k^\dagger$) is the boson annihilation
    (creation)} operator of the mode $k=1,2$, with frequency
$\omega_k$, and H.c. denotes Hermitian conjugate. Moreover,
$\kappa$ is the real coupling strength between the resonators.

To incorporate loss and gain in the cavities on the quantum level,
one can resort to the Scully-Lamb
ME~\cite{ScullyLambBook,YamamotoBook}, which has the 
form
\begin{eqnarray}\label{MES}  
\frac{d}{dt}\hat\rho&=&\frac{1}{i\hbar}\left[\hat H,\hat\rho\right] +\Big[\frac{A}{2}(\ap\hat\rho\am-\am\ap\hat\rho) \nonumber \\
&&+\frac{B}{8}\left[\hat\rho(\am\ap)^{2}+3\am\ap\hat\rho\am\ap-4\ap\hat\rho\am\ap\am\right] \nonumber \\
&&+\sum\limits_{i=1}^2\frac{\Gamma_i}{2}(\hat a_i\hat\rho\hat
a_i^{\dagger}-\hat a_i^{\dagger}\hat a_i\hat\rho) + \rm {H.
    c.}\Big],
\end{eqnarray}
given in terms of the gain $A$ and  gain saturation $B$
coefficients for the field in the active cavity. This equation
describes the dynamics of the photonic part of a quantum laser,
and, accordingly, the coefficients can be expressed as
\begin{equation}\label{AB}  
A=\frac{2g^2r}{Y^2}, \quad \text{and} \quad B=\frac{4g^2}{Y^2}A,
\end{equation}
where the parameter $g$ stands for the coupling strength between
the atoms of the gain medium and the optical field in the active
cavity, $Y$ is the decay rate of the atoms, and $r$ accounts for
the pump rate of the gain medium.  In \eqref{MES}, the total decay
rates for both cavities are given by ($i=1,2$)
\begin{eqnarray}  
\Gamma_i=C_i+\gamma_i,
\end{eqnarray}
where $C_i$ is the intrinsic loss of the $i$th cavity, and
$\gamma_i$ stands for the loss due to the possible coupling of the
$i$th cavity to the $i$th waveguide.

\subsection{Liouvillian and effective non-Hermitian Hamiltonian for the system of coupled active and passive cavities in the weak-gain-saturation regime}

The ME, given in \eqref{MES}, can be {recast} as an equation with
a Lindblad Liouvillian superoperator $\cal L$ as~\cite{Gea1998},
\begin{eqnarray}\label{Lindblad}  
\frac{d}{dt}\hat\rho&=&{\cal L}\hat\rho(t) \nonumber \\
&=&\frac{1}{i\hbar}\left[\hat H,\hat\rho\right]
-\frac{1}{2}\sum\limits_{i=1}^4\left(\hat L_i^{\dagger}\hat
L_i\hat\rho+\hat\rho\hat L_i^{\dagger}\hat L_i-2\hat
L_i\hat\rho\hat L_i^{\dagger}\right), \nonumber\\
\end{eqnarray}
where the Lindblad operators $\hat L_i$ (for $i=1,\dots,4$) are
defined as:
\begin{eqnarray}  
&\hat L_1 = \sqrt{A}\ap\left(1-\frac{B}{2A}\am\ap\right), \quad \hat L_2 = \frac{1}{2}\sqrt{3B}\am\ap, &\nonumber  \\
&\hat L_3 = \sqrt{\Gamma_1}\am, \quad \hat L_4 =
\sqrt{\Gamma_2}\hat a_2.&
\end{eqnarray}
The Lindblad form in \eqref{Lindblad} is equivalent to the ME
in Eq.~(\ref{MES}) if the terms of second order in $B\am\ap/(2A)$
are neglected in Eq.~(\ref{Lindblad}), which holds true for the
weak-gain-saturation regime.

When the active cavity is far below the lasing threshold and it is
not driven by an intense coherent field, the gain saturation
parameter $B$ can be safely dropped, and the  ME in
Eq.~(\ref{Lindblad}) reduces to the following ME with a linear
gain:
\begin{eqnarray}\label{MESR}  
\frac{d}{dt}\hat\rho&=&{\cal L}\hat\rho(t)=\frac{1}{i\hbar}\left[\hat H,\hat\rho\right] +\frac{A}{2}(2\ap\hat\rho\am-\am\ap\hat\rho-\hat\rho\am\ap) \nonumber \\
&&+\sum\limits_{i=1}^2\frac{\Gamma_i}{2}(2\hat a_i\hat\rho\hat
a_i^{\dagger}-\hat a_i^{\dagger}\hat a_i\hat\rho-\hat\rho\hat
a_i^{\dagger}\hat a_i).
\end{eqnarray}
From now on, we will always assume that the system of the coupled
active and passive cavities is linear.  Thus, we only consider
the linear ME given in Eq.~(\ref{MESR}). The ME, in
Eq.~(\ref{MESR}), as well as in Eq.~(\ref{MES}), incorporates both
the quantum jump term $\hat O\hat\rho\hat O^{\dagger}$, and the
continuous amplification or dissipation terms $\hat O\hat
O^{\dagger}\hat\rho+\hat\rho\hat O\hat O^{\dagger}$.

We notice that the Liouvillian in \eqref{MESR} is quite general,
and is {\it not} only limited to the description of quantum lasers
in the linear-gain approximation. Indeed, \eqref{MESR} describes
also an incoherently driven bosonic dimer. Recently, several
incoherent driving mechanisms were proposed
\cite{HoffmanPRL11,LebreuillyPRA17,LEBREUILLY2016836}, and the
presence of photon-photon interaction was shown to induce a
critical behavior in lattices of resonators
\cite{BiellaPRA17,ScarlatellaPRB19}. Since LEPs suggest the
presence of a dissipative phase transition \cite{Minganti2018} and
can occur also far from the thermodynamic limit, the study of the
EPs in the dimer model relates to criticality and
spontaneous-symmetry breaking characterizing the phase transition
of the full lattice model.

On the other hand, in the vast literature devoted to $\cal
PT$-symmetric systems with balanced gain and losses,  one can
often encounter the use of the phenomenological {\it
effective}  NHH:
\begin{equation}\label{Heff} 
\hat H_{\rm eff}=\hat H+\frac{i\hbar}{2}
A\ap\am-\frac{i\hbar}{2}\sum\limits_{j=1}^2\Gamma_j\hat
a^{\dagger}_j\hat a_j,
\end{equation}
with the unitary Hamiltonian $\hat H$ given in Eq.~(\ref{H}).
As one can see, this NHH incorporates the gain and loss rates as
the imaginary part of the field frequencies.

The NHH $\Hef$, in Eq.~(\ref{Heff}), gives the same dynamics for
the fields $\hat{a}_{j}$, $j=1, \,2$, as the ME in
Eq.~(\ref{MESR}), but fails to explicitly incorporate quantum
noise;  thus, making the NHH usable, in general, only in the
semiclassical limit.
A detailed
discussion of the actual semiclassical limit in this model will be
in Sec.~III.

Below, we calculate the HEPs and LEPs of the NHH $\Hef$ in
Eq.~(\ref{Heff}) and Liouvillian $\cal L$ in Eq.~(\ref{MESR}),
respectively, in both semiclassical and quantum regimes for a
given linear system in order to reveal their differences.

\subsection{Liouvillian spectrum and exceptional points}

Before we analyze the EPs of the Scully-Lamb model, let us first
briefly recall some key properties of the Liouvillian
spectrum~\cite{Minganti2019,Minganti2018}.

\subsubsection{Diagonalization of the Liouvillian superoperator}
The spectrum of the Liouvillian $\cal{L}$, given in
Eq.~(\ref{MESR}), is found according to the formula
\begin{equation}\label{LS}  
{\cal{L}}\hat\rho_i=\lambda_i\hat\rho_i,
\end{equation}
where $\lambda_i$ and $\hat\rho_i$ are the eigenvalues and
eigenmatrices of the Liouvillian, respectively. We can always order the eigenvalues and eigenmatrices in such a way that ${\rm Re}[\lambda_0]> {\rm Re}[\lambda_1]\geq {\rm Re}[\lambda_2]\geq \dots$. Moreover, since
the superoperator $\cal L$ is not necessarily Hermitian, it can acquire
both right (${\cal{L}}\hat\rho_i=\lambda\hat\rho_i$) and left
(${\cal{L}}^{\dagger}\hat\sigma_i=\lambda_i^*\hat\sigma_i$)
eigenmatrices, respectively. The left and right eigenmatrices obey
the relation ${\rm
Tr}[\hat\rho_i\hat\sigma_j]=\delta_{ij}$. If $\cal L$ is
diagonalizable, the density matrix $\hat\rho(t)$ of the system can
be written as follows
\begin{equation}\label{rho_diag}  
\hat\rho(t)=\sum\limits_ic_i(t)\hat\rho_i,
\end{equation}
where $c_i(t)=\exp\left(\lambda_it\right){\rm
Tr}[\hat\sigma_i\hat\rho(0)]$.

The eigenvalue $\lambda_0=0$ of the Liouvillian $\cal L$ in
Eq.~(\ref{LS}) defines the steady-state density matrix
$\hat\rho_{\rm ss}\propto\hat\rho_0$ of the system. The
proportionality factor depends on the normalization choice which
is done on $\hat \rho_0$. Indeed, one often induces the standard
Hilbert-Schmidt norm, so that $\|\hat \rho_0\|^2 = {\rm Tr}
\left[\hat\rho_0^\dagger \hat\rho_0\right]=1$, while instead ${\rm
Tr} \left[\hat\rho_{\rm ss}\right]=1$. For the remaining nonzero
eigenvalues $\lambda_i\neq0$ the corresponding eigenmatrices
$\hat\rho_i$ are traceless, i.e., ${\rm Tr}[\hat\rho_i]=0$.

If $\lambda_i\in{\mathbb R}$, then the corresponding eigenmatrix
$\hat\rho_i$ is Hermitian. In this case, by diagonalizing the
eigenmatrix
\begin{equation} 
\rho_i=\sum\limits_n
p^{(i)}_n|\psi_n^{(i)}\rangle\langle\psi_n^{(i)}|,
\end{equation}
 one can consider the following decomposition $\hat\rho_i=\hat\rho_i^{+}-\hat\rho_i^{-}$, where
\begin{equation} 
\hat\rho_i^{+}=\sum\limits_{n\leq \bar n}
p^{(i)}_n|\psi_n^{(i)}\rangle\langle\psi_n^{(i)}|, \quad
\text{with} \quad  p_n^{(i)}\geq0,
\end{equation}
and
\begin{equation} 
\hat\rho_i^{-}=-\sum\limits_{n> \bar n}
p^{(i)}_n|\psi_n^{(i)}\rangle\langle\psi_n^{(i)}|, \quad
\text{with} \quad p_n^{(i)}<0,
\end{equation}
and such that ${\rm Tr}[\hat\rho_i^{+}]={\rm
Tr}[\hat\rho_i^{-}]=1$. The latter stems from the fact that the
eigenmatrix $\hat\rho_i$ is traceless and one can always rearrange
the coefficients $p_n^{(i)}$ such that $p_n^{(i)}>0$ when $n\leq
\bar n$, and $p_n^{(i)}<0$ when $n> \bar n$. Now with such a
decomposition, the wave-functions constituting both
$\hat\rho_i^{\pm}$ can be compared with those comprising the
corresponding effective NHH.

When $\lambda_i\in{\mathbb C}$, the eigenmatrix $\hat\rho_i$
becomes non-Hermitian. Clearly, in this case, in order to ensure
Hermiticity of the total density matrix $\hat\rho(t)$ one has to
consider the Hermitian  symmetric $\hat\rho_i^{\rm
s}=\hat\rho_i+\hat\rho_i^{\dagger}$ and antisymmetric
$\hat\rho_i^{\rm a}=i\left(\hat\rho_i-\hat\rho_i^{\dagger}\right)$
combinations. Again, by performing the same decomposition
procedure as above, one arrives at the density matrices
\begin{equation} 
\hat\rho_i^{\rm s}=\hat\rho_i^{{\rm s}+}-\hat\rho_i^{{\rm s}-},
\quad \text{and} \quad \hat\rho_i^{\rm a}=\hat\rho_i^{{\rm
a}+}-\hat\rho_i^{{\rm a}-}.
\end{equation}
In this formalism, a Liouvillian exceptional point  is
{the point of the parameter space where two eigenmatrices of
the Liouvillian coalesce.} Since LEPs are associated with a
non-diagonalizable Liouvillian, at the critical point one has a
Jordan canonical form. With an  LEP of order 2, one has an
eigenvalue $\lambda_{\rm EP}$ and a generalized eigenmatrix
$\hat\rho'_{\rm EP}$. Consequently, \eqref{rho_diag} becomes:
\begin{equation}\label{rho_diag_EP}  
\hat\rho(t)=\sum\limits_i c_i(t)\hat\rho_i + c_{\rm
EP}(t)\hat\rho_{\rm EP}+ c'_{\rm EP}(t)\hat\rho'_{\rm EP},
\end{equation}
where
\begin{equation*}
c_{\rm EP}(t)= \exp\left(\lambda_{\rm EP}t\right){\rm
Tr}[\hat\sigma_{\rm EP}\hat\rho(0)],
\end{equation*}
while
\begin{equation*}
c'_{\rm EP}(t)= t \exp\left(\lambda_{\rm EP}t\right){\rm
Tr}[\hat\sigma'_{\rm EP}\hat\rho(0)].
\end{equation*}
Moreover, LEPs should be understood as purely dynamical phenomena.
In this Lindblad ME formalism, LEPs can emerge only for those
eigenstates of the Liouvillian with a negative real part, i.e.,
those describing the evolution of an initial density matrix
towards its steady state (for more detailed discussions, see
Refs.~\cite{Minganti2018,Minganti2019,Albert2014}).

\subsubsection{Two-time correlation functions}

A direct diagonalization of the Liouvillian necessary to access
its spectrum, however, is often extremely challenging; especially,
considering the exponentially diverging size of the Hilbert space
of the system. A TTCF could
capture the nature of EPs: A generic operator $\hat{O}$, which
does not commute with the Hamiltonian, projects the system {\it
out} of its steady state. This new density matrix is the
superposition of several Liouvillian eigenmatrices, in principle
including those associated with a LEP. For example, this idea was
used in Ref.~\cite{FinkNatPhys18} to explicitly access the
Liouvillian gap, i.e., the $\lambda_{i}$ with the smallest real part,
of a Kerr resonator. This implies that the conditional dynamics,
which follows the application of the operator $\hat{O}$, bears a
signature of the EP presence. Indeed, any TTCF can be written
as~\cite{walls_milburn_2011}
\begin{equation} 
\langle \hat{A}(t) \hat{B}(t+\tau)\rangle={\rm Tr} \left\{ \hat
A(0) e^{\mathcal{L} \tau} \left[ \hat\rho(t)\hat B(0)\right]
\right\},
\end{equation}
where the square brackets indicate that the action of the
exponential Liouvillian map must be taken on the matrix
$\rho(t)\hat B(0)$. In this regard, for the steady state we
define
\begin{equation} 
\langle \hat{A}(0) \hat{B}(\tau)\rangle_{\rm ss}={\rm Tr}
\left\{\hat A(0) e^{\mathcal{L} \tau} \left[\hat\rho_{\rm ss}\hat
B(0)\right] \right\}.
\end{equation}
The matrix $\hat \rho_{\rm ss}\hat B$ is, in general, different
from $\hat\rho_{\rm ss}$. Therefore, we can express it in terms of the
generalized eigenmatrices $\hat{\rho}_i$ of the Liouvillian (
including $\hat\rho'_{\rm EP}$), that is,
\begin{equation}  
\hat{\rho}_{\rm ss}\hat B = \sum_i c_i \hat{\rho}_i.
\end{equation}
Because we have used the spectral decomposition of the
Liouvillian, by recalling the linearity of the trace, we have
\begin{equation} \label{TTCFdef} 
\langle \hat{A}(0) \hat{B}(\tau)\rangle_{\rm ss}=\sum_i c_i  {\rm
Tr} \left\{\hat A(0) e^{\mathcal{L} \tau} \left[\hat{\rho}_i
\right] \right\}.
\end{equation}

We have two possible cases: (i) For a system without   EPs or
away from  them, the Eq.~(\ref{TTCFdef}) reads
\begin{equation}  
\langle \hat{A}(0) \hat{B}(\tau)\rangle_{\rm ss}=\sum_i c_i
e^{\lambda_i \tau}  {\rm Tr} \left\{\hat A(0)  \hat{\rho}_i
\right\}.
\end{equation}
Indeed, for long times, only the slowly decaying fields are
relevant and
\begin{equation}\label{Mil_TTCF}  
\langle \hat{A}(0) \hat{B}(\tau)\rangle_{\rm ss}\simeq c_0 {\rm
Tr} \left\{A  \hat{\rho}_0 \right\} + c_1 e^{\lambda_1 \tau}  {\rm
Tr} \left\{A  \hat{\rho}_1 \right\}+ \ldots
\end{equation}
In this regard, $\langle \hat{A}(0) \hat{B}(\tau)\rangle_{\rm ss}$
as a function of time $\tau$ describes an exponential decay
towards the steady-state value $c_0 {\rm Tr} \left\{\hat A(0)
\hat{\rho}_0 \right\}$.
\\
(ii) In the presence of an LEP, one has
\begin{equation} 
\langle \hat{A}(0) \hat{B}(\tau)\rangle_{\rm ss}=\sum_i c_i
\tau^{n_i} e^{\lambda_i \tau}  {\rm Tr} \left\{\hat A(0)
\hat{\rho}_i \right\},
\end{equation}
where $n_i$ is the degree of  degeneracy of the EP associated
with the eigenmatrix $\hat{\rho}_i$. {For example, for an EP of
degree 3, we would have a contribution of
\begin{equation}\label{eq23} 
\begin{split}
 e^{\lambda_i \tau} &\left[c_i  {\rm Tr} \left\{\hat A(0)  \hat{\rho}_i \right\} + c_{i+1} \tau {\rm Tr} \left\{\hat A(0)  \hat{\rho}_{i+1}\right\} \right.\\& \left.   \quad + c_{i+2} \tau^2  {\rm Tr} \left\{\hat A(0)  \hat{\rho}_{i+2} \right\}\right],
\end{split}
\end{equation}}
in the expansion of Eq.~(\ref{TTCFdef}).

In this regard, a deviation from an exponential decay signals the
presence of an EP. This implies that the conditional dynamics,
which follows the application of the operator $\hat{O}$, bears a
signature of the presence of an EP.

\section{Hamiltonian and Liouvillian exceptional points  in the Semiclassical regime}

Here we study the EPs of both the non-Hermitian Hamiltonian and
Liouvillian in the semiclassical limit. Hence, we consider
the two-cavity system, shown in Fig.~\ref{fig1}, populated by many
photons $\langle \hat n\rangle\gg1$, i.e., the system can be
probed by intense coherent fields.  Such an assumption does not
allow us to represent the Liouvillians in their matrix form, due to
the rapidly exponentially diverging size of the latter. The
weak-gain case, where the Liouvillian can be exactly diagonalized,
will be investigated in Sec.~IV. Here we resort rather
to the two-mode formalism to deduce the presence of an LEP.

{We note that the effective Hamiltonian, 
studied here, describes the gain and loss as the imaginary parts
of the frequencies of quantum fields [see Eq.~(\ref{Heff})]. Such a
Hamiltonian arises from the mean-field approximation and, as a
result, its use is justified in the semiclassical regime, when
considering intense coherent  fields. The NHH associated with this
model explicitly exhibits a $U(1)$ Hamiltonian symmetry, implying
that the subspaces corresponding to different numbers of excitations do not mix,
even if the total number of excitations is not conserved. On the
other hand, this symmetry is broken in the corresponding Liouvillian because of
the presence of the quantum-jump terms. The Liouvillian approach describes a mixed-state dynamics obtained
by averaging over many pure-state quantum trajectories, where
quantum jumps induce transitions between manifolds corresponding
to different numbers of excitations.
Nonetheless, in the semiclassical limit with many excitations, the
action of the creation and annihilation operators, associated with
a quantum jump, scales as $\sqrt{n}$ in a cavity with $n$
excitations, while the other energy terms scale as $n$. Therefore,
adding or removing a single excitation does not drastically change
typical properties of the system even at the level of its
eigenvectors. As a result, in the frequency spectrum, one might
expect some similarity between an NHH and the corresponding
Liouvillian in the semiclassical limit.}

\subsection{Hamiltonian exceptional points}

Let us first find an EP of the effective NHH $\Hef$, in
Eq.~(\ref{Heff}).

By introducing the operator  vector $\hat a=\left(\hat a_1, \hat
a_2\right)^T$,  one can recast the NHH $\Hef$, in Eq.~(\ref{Heff}),
in the matrix form as
\begin{equation}\label{Hef_M}  
\Hef = \hat a^{\dagger}H\hat a, \quad \text{where}\quad
H=\begin{pmatrix}
\omega_c+i\frac{A-\Gamma_1}{2} & -i\kappa \\
i\kappa & \omega_c-i\frac{\Gamma_2}{2}
\end{pmatrix},
\end{equation}
From  Eq.~(\ref{Hef_M}), one then can immediately find the
eigenvalues of the Hamiltonian $\Hef$,
\begin{equation}\label{SHeff_eig}  
\nu_{1,2}=\omega_c+\frac{i}{4}\left(A-\Gamma_+\right)\pm
\frac{i}{4}\beta,
\end{equation}
where $\beta=\sqrt{(A-\Gamma_-)^2-16\kappa^2}$, and
$\Gamma_{\pm}=\Gamma_1\pm\Gamma_2$. 

{The complex eigenvalues $\nu_i$ indicate the non-Hermitian character of the Hamiltonian $\Hef$.
Moreover, because of this non-Hermiticity, the operator $\Hef$ can attain both right $|\psi\rangle$ and left $\langle\tilde\psi|$ eigenvectors via relations
\begin{equation}  
 \Hef|\psi_i\rangle=\nu_i |\psi_i\rangle \quad \text{and} \quad  \langle\tilde\psi_i|\Hef=\nu_i\langle\tilde\psi_i|, 
\end{equation} 
respectively. Hereinafter, without loss of generality, we consider only  right eigenvectors $|\psi_i\rangle$ of the NHH $\Hef$, since the HEPs are defined equivalently using either set of vectors.}

The corresponding right eigenvectors
become
\begin{equation}\label{SHeff_eigv}  
|\psi_{1,2}\rangle=\frac{1}{N_{1,2}}\begin{pmatrix}
{A-\Gamma_-\pm\beta} \\
4\kappa
\end{pmatrix},
\end{equation}
where $N_\pm$ is the corresponding normalization coefficient.

By analyzing Eqs.~(\ref{SHeff_eig}) and~(\ref{SHeff_eigv}), one
comes to the conclusion that, in the semiclassical regime, the NHH
$\Hef$ has an HEP, where both  eigenvalues and eigenvectors
coalesce when
\begin{equation}\label{sHEP}   
\kappa_{\rm HEP}^{\rm s}=\frac{1}{4}\left|A-\Gamma_-\right|.
\end{equation}

{At the HEP, the two linearly independent eigenvectors $|\psi_{1,2}\rangle$ coalesce to a single eigenvector 
\begin{equation} 
|\psi_{\rm HEP}\rangle\equiv\begin{pmatrix}
1 \\
1
\end{pmatrix}.
\end{equation}
 In this case, 
the $2\times2$ NHH $\Hef$ becomes nondiagonazible, thus acquiring a Jordan form. This means that, at the HEP, the generalized eigenspace of the NHH $\Hef$ is spanned by the vector $|\psi_{\rm HEP}\rangle$ and a pseudo-eigenvector $|\psi_{\rm HEP}'\rangle$, which is obtained from $|\psi_{\rm HEP}\rangle$ via a Jordan chain relation and reads
\begin{equation} 
|\psi'_{\rm HEP}\rangle\equiv\begin{pmatrix}
-1 \\
1
\end{pmatrix}.
\end{equation}
 For details regarding pseudo-eigenvectors see, e.g., Refs.~\cite{Hashimoto2015,Kanki2017}. }

It is also worth noting that the NHH $\Hef$ in Eq.~(\ref{Heff}) fails
to incorporate spontaneous emission, since $\Hef\ket 0=0$.
Obviously, because of the presence of the gain process in the
active cavity, the probability of spontaneous emission is nonzero.
To overcome this difficulty, one can apply the
Heisenberg equations to the quantum field operators $\hat a_j$ ($j=1,2$),
\begin{equation*}
\frac{{\rm d}\hat{a}_j}{{\rm d}t}=\frac{1}{i\hbar}[\hat a_j,\Hef],
\end{equation*}
with the phenomenologically introduced quantum Langevin
forces~\cite{ZollerBook},
\begin{eqnarray}\label{LE}  
\frac{{\rm d}}{{\rm d}t}\hat a_1&=&\frac{A-\Gamma_1}{2}\hat a_1-\kappa\hat a_2+\sqrt{A}\hat g_1^{\dagger}+\sqrt{\Gamma_1}\hat l_1, \nonumber \\
\frac{{\rm d}}{{\rm d}t}\hat a_2&=&-\frac{\Gamma_2}{2}\hat
a_2+\kappa\hat a_1+\sqrt{\Gamma_2}\hat l_2,
\end{eqnarray}
where $\hat g_j^{\dagger}$ ($\hat l_j$) is the quantum noise amplification
(dissipation) operator of the $j$th cavity, with the commutation
relations $[\hat O_j(t),\hat
O_k^{\dagger}(t')]=\delta_{jk}\delta(t-t')$, for $\hat O=\hat g,
\hat l$, $j=1,2$.

Now, the equations of motion for the quantum fields given in
Eq.~(\ref{LE}) can provide the same fields dynamics as by the
Liouvillian $\cal L$~\cite{AgarwalBook},  which we consider below.

Importantly, in order to properly describe the spectral properties
of the fields,  the rate equations in Eq.~(\ref{LE}), for the
active cavity field $\hat a_1$, should contain both  amplification
and dissipation noise operators. Otherwise, one can arrive at
wrong conclusions (see Appendix~A for details). We stress that the
omission of the dissipation noise operator in the active cavity,
in Eq.~(\ref{LE}), has become widespread in  the literature,
especially in that devoted to $\cal PT$-symmetric systems.

\subsection{Liouvillian exceptional points}

As we discussed, it is, in general, challenging to find an LEP of
the Liouvillian $\cal L$ in Eq.~(\ref{MESR}), especially in the
semiclassical regime. However, one could infer the presence of
LEPs using the TTCFs of the fields, as it was described in
Sec.~II~B. Below, we compute $\langle\hat a_j^{\dagger}(0)\hat
a_j(\tau)\rangle_{\rm ss}$ for the field in the $j$th cavity,
$j=1,2$, in the steady state, to demonstrate its ability to
capturing the EPs of the Liouvillian. We note that this method,
which enables to reveal the dynamics of the Liouvillian, can be
extended to high-order TTCFs~\cite{Haake1971}, as it was
experimentally done in, e.g., Ref.~\cite{FinkNatPhys18}. Moreover,
our calculations are made simpler by the absence of a driving
field in the \eqref{H}, i.e., the TTCF does not involve a coherent
part due to an external driving laser field, and will only capture
the incoherent part of the TTCF induced by the gain in the active
cavity. We note that, in the presence of a coherent field, the
dynamical character of the incoherent part of the TTCF would not
change qualitatively; thus, we could perform the same analysis for
that model. Finally, we stress that this method indicates the
presence of an LEP, but it does not provide neither the structure
of the eigenmatrices of the Liouvillian nor their relation to the
eigenvectors of the NHH. These two can differ substantially, as it
will be shown in the next section.

\subsubsection{Computation of the {two-time correlation function}}

To obtain the TTCF  one may invoke the quantum regression theorem,
which states that the equations of motion for  system operators
are also the equations of motion for their correlation functions.
To express this theorem mathematically, one can write the
following equation~\cite{CarmichaelBook}:
\begin{equation}\label{RF}  
\frac{d}{d\tau}\langle\hat O(t)\boldsymbol{\hat
A}(t+\tau)\rangle=\boldsymbol{M}\langle\hat O(t)\boldsymbol{\hat
A}(t+\tau)\rangle,
\end{equation}
where $\boldsymbol{\hat A}=[\hat A_1, \hat A_2, \dots, \hat
A_\nu]$ is the vector of a complete set of the  system operators
$\hat A_\mu$, in the sense that the averages $\langle\hat
A_\mu\rangle$, $\mu=1,2,\dots,\nu$,  form the  set of coupled
linear equations with the evolution matrix $\boldsymbol{M}$. The
operator $\hat O$ can be arbitrary, not necessarily belonging to
$\hat A_\mu$.

For the studied system of coupled active and passive cavities,
governed by a ME with the Liouvillian $\cal L$ in
Eq.~(\ref{MESR}), and with the Hamiltonian in Eq.~(\ref{H}), the
complete set is formed by the following vector $\boldsymbol{\hat
A}=[\hat a_1,\hat a_2]$ of the field operators $\hat a_1,\hat
a_2$. The evolution matrix $\boldsymbol{M}$ is found to be
\begin{equation}\label{M}  
\boldsymbol{M}=-iH,
\end{equation}
where $H$ is given in Eq.~(\ref{Hef_M}).

Now, by combining Eqs.~(\ref{RF}) and~(\ref{M}), and using the
operators $\hat a_j^{\dagger}$, $j=1,2$ instead of the operator
$\hat O$, one obtains the following solution for the TTCF,
\begin{eqnarray}\label{TTCF}  
&\begin{pmatrix}
\langle\hat a_j^{\dagger}(t)\hat a_j(t+\tau)\rangle \\
\langle\hat a_j^{\dagger}(t)\hat a_k(t+\tau)\rangle
\end{pmatrix}=\exp\left(\boldsymbol{M}\tau\right)
\begin{pmatrix}
\langle\hat a_j^{\dagger}(t)\hat a_j(t)\rangle \\
\langle\hat a_j^{\dagger}(t)\hat a_k(t)\rangle
\end{pmatrix},&
\end{eqnarray}
for $j,k=1,2,$ $ j\neq k$.

The TTCF in the steady state can be obtained by sending
$t\rightarrow\infty$ in Eq.~(\ref{TTCF}). As Eq.~(\ref{TTCF})
indicates, in order to find correlation functions, one needs first
to know the averages of the photon numbers in each cavity as well
as the averages  $\langle\hat a_j^{\dagger}(t)\hat a_k(t)\rangle$.

Again, by applying the master equation in Eq.~(\ref{MESR}) to the
operators $\hat a_j^{\dagger}\hat a_k$ and $\hat a_j^{\dagger}\hat
a_k$, one obtains their averages in the steady state as follows:
\begin{eqnarray}\label{NSS}  
\langle\hat a_1^{\dagger}\hat a_1\rangle_{\rm ss} &=& {A\left(4\kappa^2-G_1\Gamma_2+\Gamma_2^2\right)}f, \nonumber \\
\langle\hat a_2^{\dagger}\hat a_2\rangle_{\rm ss} &=& {4\kappa^2A}f, \nonumber \\
\langle\hat a_1^{\dagger}\hat a_2\rangle_{\rm ss} &=& \langle\hat
a_2^{\dagger}\hat a_1\rangle_{\rm ss}={2\kappa\Gamma_2A}f,
\end{eqnarray}
where $G_1=A-\Gamma_1$ represents the total net gain in the active
cavity, and
$f^{-1}={\left(4\kappa^2-G_1\Gamma_2\right)(\Gamma_2-G_1)}$ is a
normalization factor.

\begin{figure}[tb] 
\includegraphics[width=0.33\textwidth]{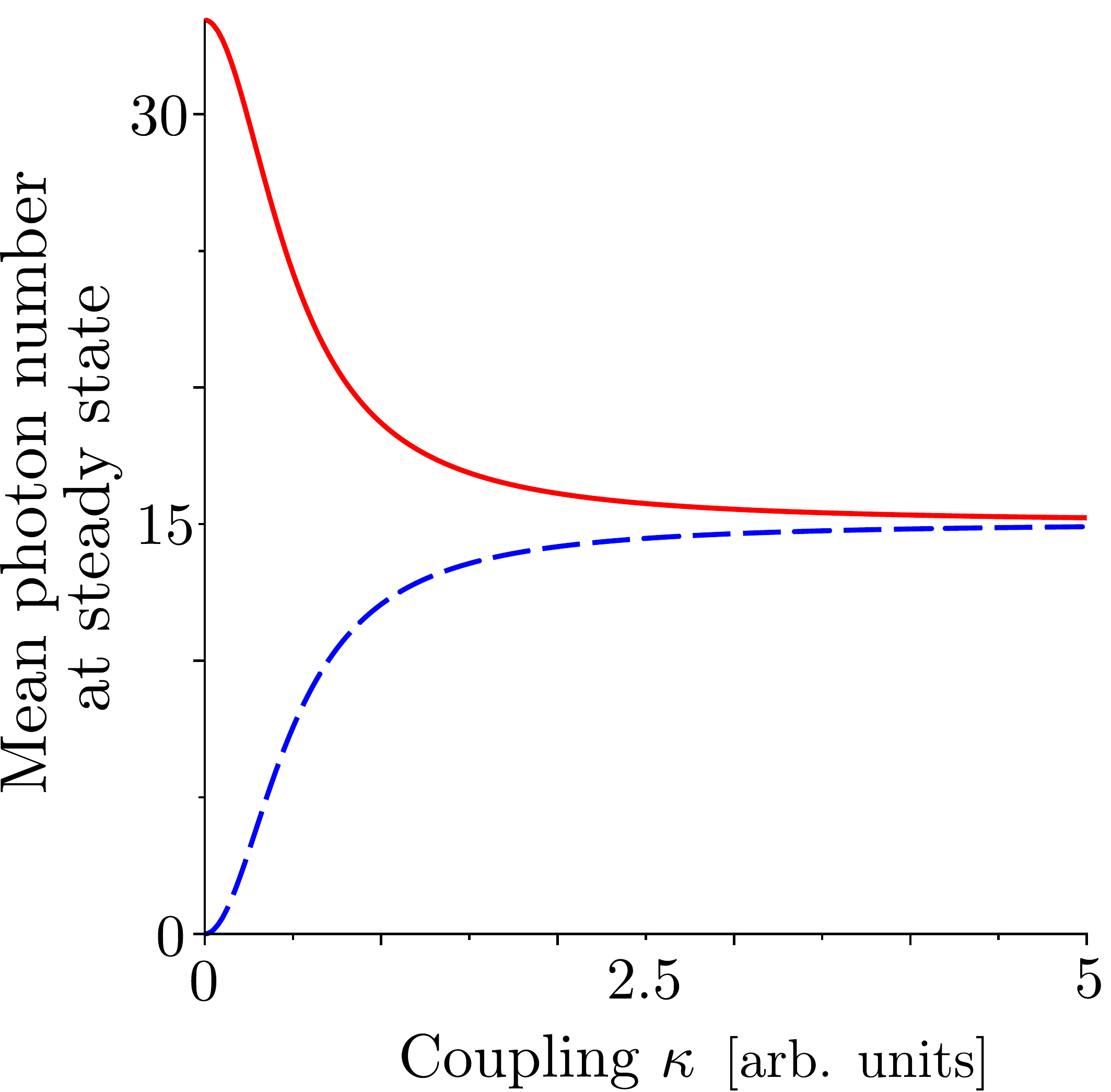}
\caption{Mean photon-numbers $\langle\hat n_1\rangle$ in the
active cavity (red solid curve) and $\langle\hat n_2\rangle$ in
the passive cavity (blue dashed curve) in the steady state as a
function of the intercavity coupling $\kappa$. The intrinsic gain
and loss are balanced in the system, i.e., to satisfy the
condition $A-C_1-C_2=0$ , where the gain
$A=30.1$ [arb.~units], the intrinsic loss in the passive cavity
$C_2=0.1$ [arb.~units], and the coupling of both cavities to the
waveguides as $\gamma=1$ [arb.~units] (see Fig.~\ref{fig1}).
}\label{fig2}
\end{figure}
As an example, in Fig.~\ref{fig2} we plot the averages of the
photon numbers in the steady state in both cavities given in
Eq.~(\ref{NSS}), as a function of the intercavity coupling
strength $\kappa$. The system is chosen to balance
\emph{intrinsic} gain and losses , i.e., one imposes the condition
$A-C_1-C_2=0$ simulating the {\it effective} $\cal PT$-symmetric
regime~\cite{Liang2017}. Such a symmetry is called {\it effective}
since  the \emph{total} gain and losses are not balanced due to
nonzero waveguide coupling $\gamma\neq0$;  thus, breaking the
genuine $\cal PT$-symmetry (for details see also
Ref.~\cite{Arkhipov2019b}). As Fig.~\ref{fig2} indicates, the
average steady-state number of photons in both cavities can be
large, due to the interplay between spontaneous emission and the
gain in the active cavity [c.f. \eqref{NSS}]. By varying the
coupling strength $\kappa$ between the cavities, one obtains
different values of the photon numbers in the resonators, which
become identical in the limit $\kappa\rightarrow\infty$  (see
Fig.~\ref{fig2}):
\begin{equation*}
\langle\hat n_1\rangle=\langle\hat
n_2\rangle=\frac{A}{\Gamma_+-A}.
\end{equation*}
Photon number fluctuations are large too. For instance, for
$\kappa=0$, the dispersion of the number of photons in the active
cavity becomes $\sigma(\langle\hat
n\rangle)=\sqrt{\Gamma_1/A}\langle\hat n\rangle$, which indicates
the thermal character of the gain.

\begin{figure}[tb] 
\includegraphics[width=0.48\textwidth]{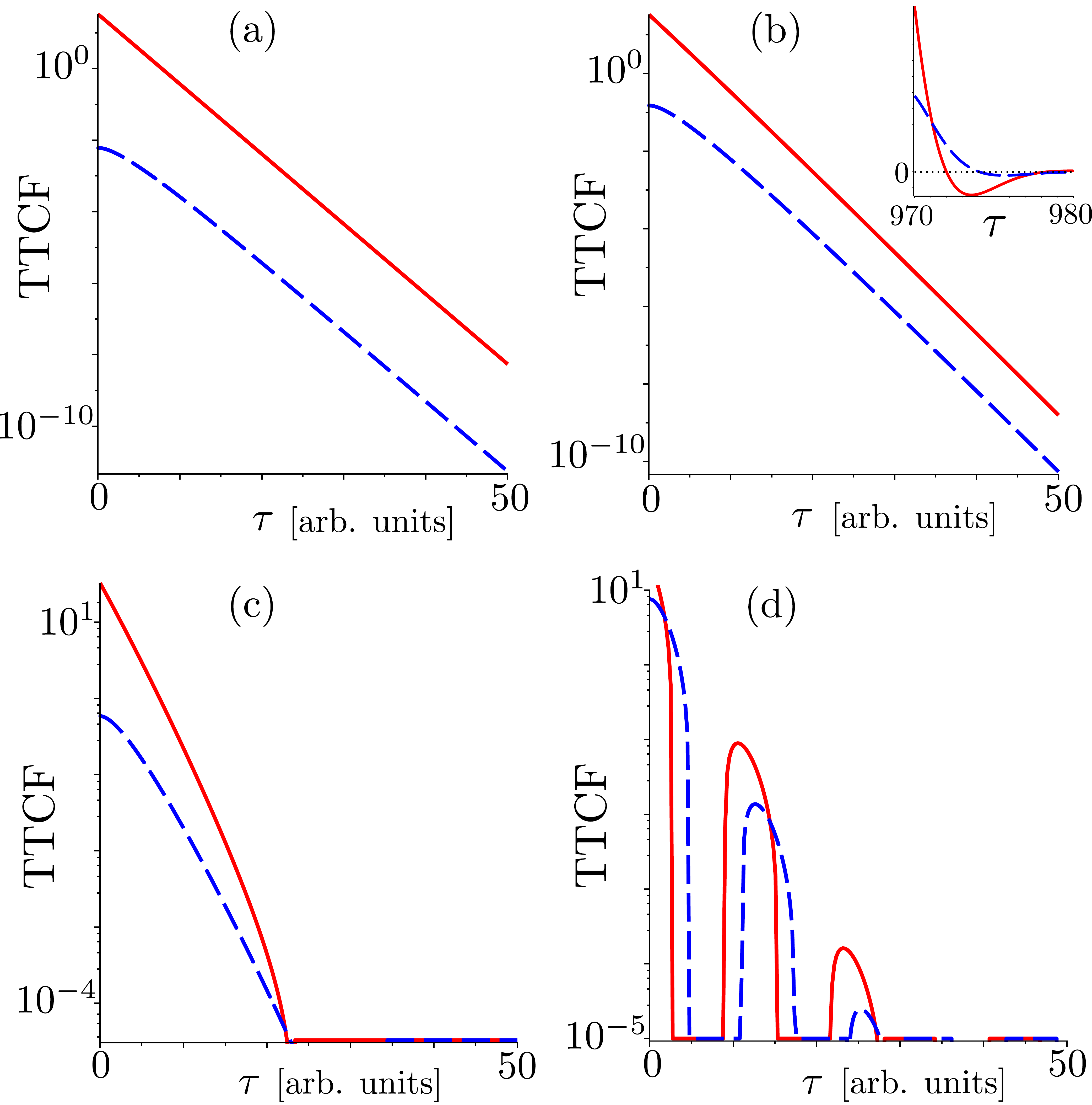}
\caption{ Two-time correlation function $\langle\hat
a_j^{\dagger}(0)\hat a_j(\tau)\rangle_{\rm ss}$ in the steady
state, according to Eq.~(\ref{TTCFf}), in the rotating reference
frame $\omega_c$, for the active (red solid curve) and passive
(blue dashed curve) cavities, for different values of the
intercavity coupling $\kappa$: (a) $\kappa=0.01$ [arb.~units], (b)
$\kappa=0.0501$ [arb.~units], (c) $\kappa=0.1$ [arb.~units], and (d)
$\kappa=0.5$ [arb. units]. The other system parameters are the same as
in Fig.~\ref{fig2}. {For  this system, the Liouvillian EP is found
at $\kappa=0.05$ [arb. units], according to Eq.~(\ref{TTCFf}), i.e., the point
at and above which the TTCF  fails to  demonstrate solely an
exponential decay [see the inset in panel (b)]. In order to
capture the deviation of the TTCF from the explicit exponential
behaviour right above  the EP, one  might need longer correlation
times $\tau$ [see the inset in panel (b)]. All panels are shown in
a logarithmic scale, except the inset in panel (b). }}\label{fig3}
\end{figure}

Now, combining together Eqs.~(\ref{TTCF}) and~(\ref{NSS}), one
arrives at the formula for the TTCF in both cavities in the steady
state, {\it and away from the LEP}, which writes:
\begin{eqnarray}\label{TTCFf}  
\langle\hat a_1^{\dagger}(0)\hat a_1(\tau)\rangle_{\rm ss}&=&u_2\exp(-i\nu_1\tau)+u_1\exp(-i\nu_2\tau), \nonumber \\
\langle\hat a_2^{\dagger}(0)\hat a_2(\tau)\rangle_{\rm ss}&=&v_2\exp(-i\nu_1\tau)+v_1\exp(-i\nu_2\tau), \nonumber \\
\end{eqnarray}
where $\nu_{1,2}$ are the eigenfrequencies of the NHH in
Eq.~(\ref{SHeff_eig}), and  $u_{1,2}$, $v_{1,2}$ are functions of
the system parameters given in Appendix~B.

Equation~(\ref{TTCFf}) implies that the dynamics of the TTCF, away
from the LEP, imposed by the Liouvillian is similar to that of the
NHH $\Hef$ imposed on the fields. By comparing Eq.~(\ref{TTCFf})
and Eq.~(\ref{Mil_TTCF}), one can see that the rate of decay of
these TTCF is exactly captured by the NHH.
 Most importantly, as it
follows from Eq.~(\ref{TTCFf}), the position of at least one of
the LEPs coincides with that of the HEP:
\begin{equation}\label{sLEP1}  
\kappa^{\rm s}_{\rm LEP}=\kappa^{\rm s}_{\rm
HEP}=\frac{1}{4}\left|A-\Gamma_-\right|.
\end{equation}
{When $\kappa<\kappa^{\rm s}_{\rm LEP}$,  the TTCFs in
Eq.~(\ref{TTCFf}) exhibit a simple exponential decay, as described
by a superposition of two exponents of the Liouvillian eigenvalues $\nu_1$ and $\nu_2$.}

{When the intercavity coupling $\kappa$ equals $\kappa^{\rm
s}_{\rm LEP}$, by considering a rotating reference frame  at the
cavity frequency $\omega_c$,  the TTCFs in Eq.~(\ref{TTCFf})
reduce to:
\begin{equation}\label{PQ} 
\langle\hat a_i^{\dagger}(0)\hat a_i(\tau)\rangle_{\rm ss}
=\exp\left(\frac{1}{4}\lambda\tau\right)(P_i+Q_i\tau), \quad i=1,2,
\end{equation}
where $\lambda=A-\Gamma_+<0$, and the values of the constants
$P_i$ and $Q_i$ are given in Appendix~B. We just note here that
the expressions for $P_{1,2}$ and $Q_2$ are always
positive-valued, whereas the values of $Q_1$ can be either
positive or negative, {depending on whether the expression
$A-\Gamma_-$ is positive or negative, respectively (see
Appendix~B, for details). Thus, for linear systems with the $\cal
PT$-symmetry, including the {\it effective} $\cal PT$-symmetry,
the coefficient $Q_1$ is always positive and becomes proportional
to the intercavity-coupling strength $\kappa$.} To experimentally
determine a LEP from the TTCFs in Eq.~(\ref{PQ}), one might need
to implement curve fitting techniques to capture the deviation of
the TTCF from a simple exponential decay, when increasing the
intercavity coupling $\kappa$. In particular,  if $A-\Gamma_-<0$,
i.e., $Q_1<0$, then a LEP can be directly defined from the arising
negative values of the TTCF in the active cavity, according to
Eq.~(\ref{PQ}).}

{On the other hand, in general, right above the EP, i.e., when
$\kappa>\kappa^{\rm s}_{\rm LEP}$, both TTCFs in Eq.~(\ref{TTCFf})
can acquire negative values due to the arising oscillatory term
$\beta$ in the rotating frame $\omega_c$. In order to catch these
arising negative values in the TTCFs, the observation of longer
coherence times might be needed  [see the inset in
Fig.~\ref{fig3}(b)]. Additionally, these oscillations make the TTCFs
substantially deviate from the simple exponential decay when
increasing $\kappa$ [see Figs.~\ref{fig3}(c)--\ref{fig3}(d)].}

\subsubsection{Power spectrum}
We note that in real experimental situations, it might be very
challenging  to measure a TTCF, necessary to determine the exact
position of the LEP. In this case, one can use complementary
frequency space analysis, where instead of the TTCF, one just
measures the power spectra of the detected fields. Those power
spectra can provide an intuitive and comprehensive interpretation
of the EP. Namely, the presence of the EP, e.g., of the second
order, can be revealed by a squared Lorentzian lineshape in the
power spectrum, corresponding to a coalescence of two resonance
peaks. The latter technique has been already successfully used in,
e.g., Ref.~\cite{Peng2014}.

The formula for the power emission spectra in the $j$th cavity
expressed via the TTCF reads
\begin{equation}\label{Sw}  
S_j(\omega)=\frac{1}{2\pi}\int\limits_{-\infty}^{\infty}\langle\hat
a^{\dagger}_j(0)\hat a_j(\tau)\rangle_{\rm ss} e^{i\omega
\tau}\rm{d}\tau.
\end{equation}

By combining Eqs.~(\ref{M})--(\ref{NSS}) and (\ref{Sw}), one
obtains the emission spectra in the active and passive cavities:
\begin{equation}\label{S1S2}  
S_1(\omega)=\frac{AF}{2\pi}\left[\Delta^2+\frac{\Gamma_2^2}{4}\right],
\quad S_2(\omega)=\frac{\kappa^2AF}{2\pi},
\end{equation}
where
\begin{eqnarray}
&F=\left[{\omega_+^2\omega_-^2+\frac{1}{4}{(G_1^2+\Gamma_2^2)}\Delta^2+\frac{1}{16}{G_1\Gamma_2(G_1\Gamma_2-8\kappa^2)}}\right]^{-1}&,
\nonumber
\end{eqnarray}
with $\Delta=\omega-\omega_c$ being the frequency detuning,
$\omega_\pm=\Delta\pm\kappa$, and the net gain in the active
cavity is $G_1=A-\Gamma_1<0$.

{Before we start the spectral-power analysis based on
Eq.~(\ref{S1S2}), first we would like to draw a small remark. Note
that because of our definition of the power spectra given in terms
of the non-Hermitian annihilation operators $\hat a$ in
Eq.~(\ref{Sw}), the spectrum of the vacuum is set to
zero~\cite{AgarwalBook,Hauer2015}. The quantum-field spectral
power $S(\omega)$ vanishes for the vacuum, as implied by \eqref{sLEP1}.
 Indeed, the spectral power is defined as the Fourier
transform of a two-time average of the non-Hermitian boson
operators $\hat a$ and $\hat a^{\dagger}$. In this case, the
spectral power becomes proportional to the mean photon number in the
steady state, which for the vacuum is zero, regardless of the
presence of the dissipation noise operators. On the other hand,
when performing a real experiment, one measures the spectrum of
the Hermitian electric field $\hat{\cal E}\equiv\hat a+\hat
a^{\dagger}$, which for the vacuum in the cavity with frequency
$\omega_0$ and loss rate $\Gamma$ gives the following nonvanishing
spectral power:
\begin{equation*}
S_{\hat{\cal E}}(\omega)\equiv\int\langle{\hat{\cal E}(0)}
{\hat{\cal E}(\tau)}\rangle\exp(i\omega \tau){\rm d}\tau=
\frac{\Gamma}{(\omega-\omega_c)^2+\frac{\Gamma^2}{4}},
\end{equation*}
where the amplitude of the vacuum fluctuations is set to 1.}

Now, by inspecting Eq.~(\ref{S1S2}), one can see that the emission
spectra in both cavities are provided mainly by the gain
 $A$. In particular, for a fixed intercavity coupling
$\kappa$,  both power spectra $S_1(\omega)\rightarrow0$ and
$S_2(\omega)\rightarrow0$, if $A\rightarrow0$.
On the other hand,
the power spectrum $S_2$ in the passive cavity is always zero,
whenever $\kappa=0$, regardless of the values of the gain $A$ in
the active cavity, as expected. Moreover, the derived formulas in
Eq.~(\ref{S1S2}) show that the emission spectra in both cavities
are, in general, squared Lorentzians~\cite{Yoo2011}. The latter confirms that the
system can experience a mode-splitting phenomenon, i.e., there is
a point in  parameter space where two resonances coalesce.

The mode splitting, i.e., the appearance of the squared
Lorentzians, occurs at different $\kappa$ for the two cavities,
and it is defined via (see Appendix~B for details):
\begin{eqnarray}\label{LEP_kappa}  
\kappa_{1}&=&\frac{\sqrt{\Gamma_2}}{2}\left[\sqrt{(G_1-\Gamma_2)^2
+\Gamma_2^2}+(G_1-\Gamma_2)\right]^{\frac{1}{2}}, \nonumber \\
\kappa_{2}&=&\frac{\sqrt{2}}{4}\sqrt{G_1^2+\Gamma_2^2}.
\end{eqnarray}
{This mode-splitting difference is due to the fact that the
system has an {\it effective} $\cal PT$-symmetry. This means that
the uncompensated losses, due to the coupling of the cavities to the
waveguides, affect the two mode resolution in both cavities at the
same value of $\kappa$. Moreover, the larger is the uncompensated
loss, the larger is the mode-splitting difference.}

A comparison of Eqs.~(\ref{sLEP1}) and~(\ref{LEP_kappa}) leads us
to the conclusion that the LEPs, which are exactly determined from
the TTCF, and those obtained via power-spectra analysis are, in
general, are different.

These spectral bifurcation points (SBPs) of power spectra, given
in Eq.~(\ref{LEP_kappa}), converge to the LEP defined from the
TTCF in Eq.~(\ref{sLEP1}) only in the limit when the total loss
and gain in the system become balanced, i.e, when
$(A-\Gamma_1-\Gamma_2)\to0$. This means that the extra losses
induced by the imbalance of the net gain and damping in the active
and passive cavities strongly affects the resolution of the
genuine LEP exploiting the power spectrum. We also remark that, in
the limit when $(A-\Gamma_1-\Gamma_2)\to0$, the active cavity
approaches the lasing threshold, where the system linearity
assumption can, in general, fail, and possibly lead to nonphysical
results. Hence, since the ``true'' LEP is captured by the TTCF,
the SBPs can be seen as an approximation of the LEP.

\begin{figure}[tb] 
\includegraphics[width=0.5\textwidth]{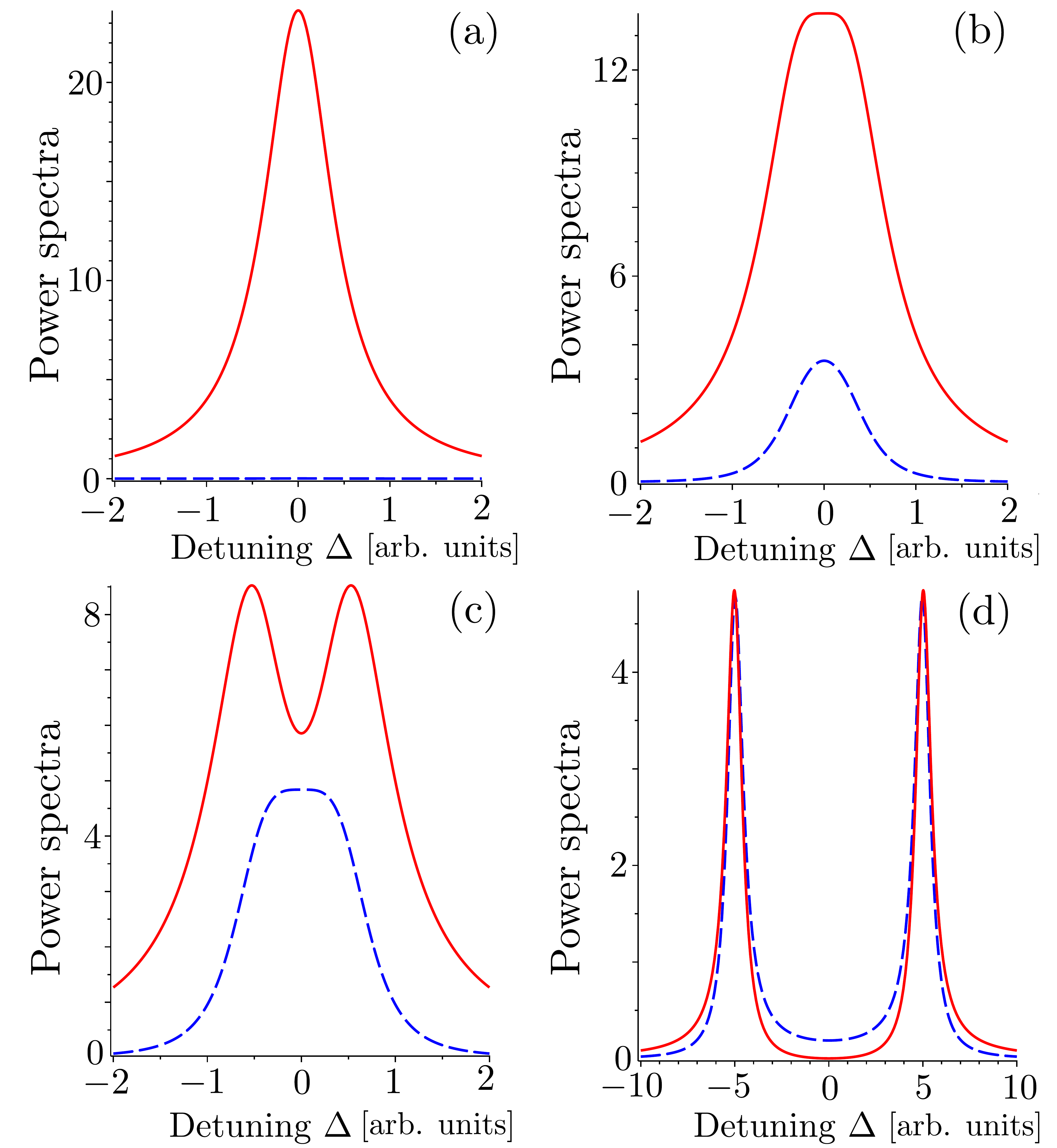}
\caption{Power spectra $S_j(\omega)$, according to
Eq.~(\ref{S1S2}),  in the active (red solid curve) and passive
(blue dashed curve) cavities versus the frequency detuning
$\Delta=\omega-\omega_c$ for different values of the intercavity
coupling $\kappa$: (a) $\kappa=0.1$ [arb.~units], (b) $\kappa=0.278$
[arb.~units], (c) $\kappa=0.5$ [arb.~units], and (d) $\kappa=5$ [arb.~units]. We
assumed that the system has an intrinsic balanced gain and losses
satisfying the relation  $A-C_1-C_2=0$. The system parameters are
the same as in Fig.~\ref{fig2}. Near  SBPs, given in
Eq.~(\ref{LEP_kappa}), the spectra exhibit squared Lorentzian
lineshapes [see panels (b)-(c)]. While far away from the SBPs, the
spectra are Lorentzian with one peak below the SBPs, and two peaks
above the SBPs [see panels (a) and (d)]. This figure demonstrates
that, in general, the Liouvillian EP can not be faithfully
determined from the power-spectra analysis, in contrast to the
TTCF, shown in Fig.~\ref{fig3}. }\label{fig4}
\end{figure}
\begin{figure}[tb] 
\includegraphics[width=0.41\textwidth]{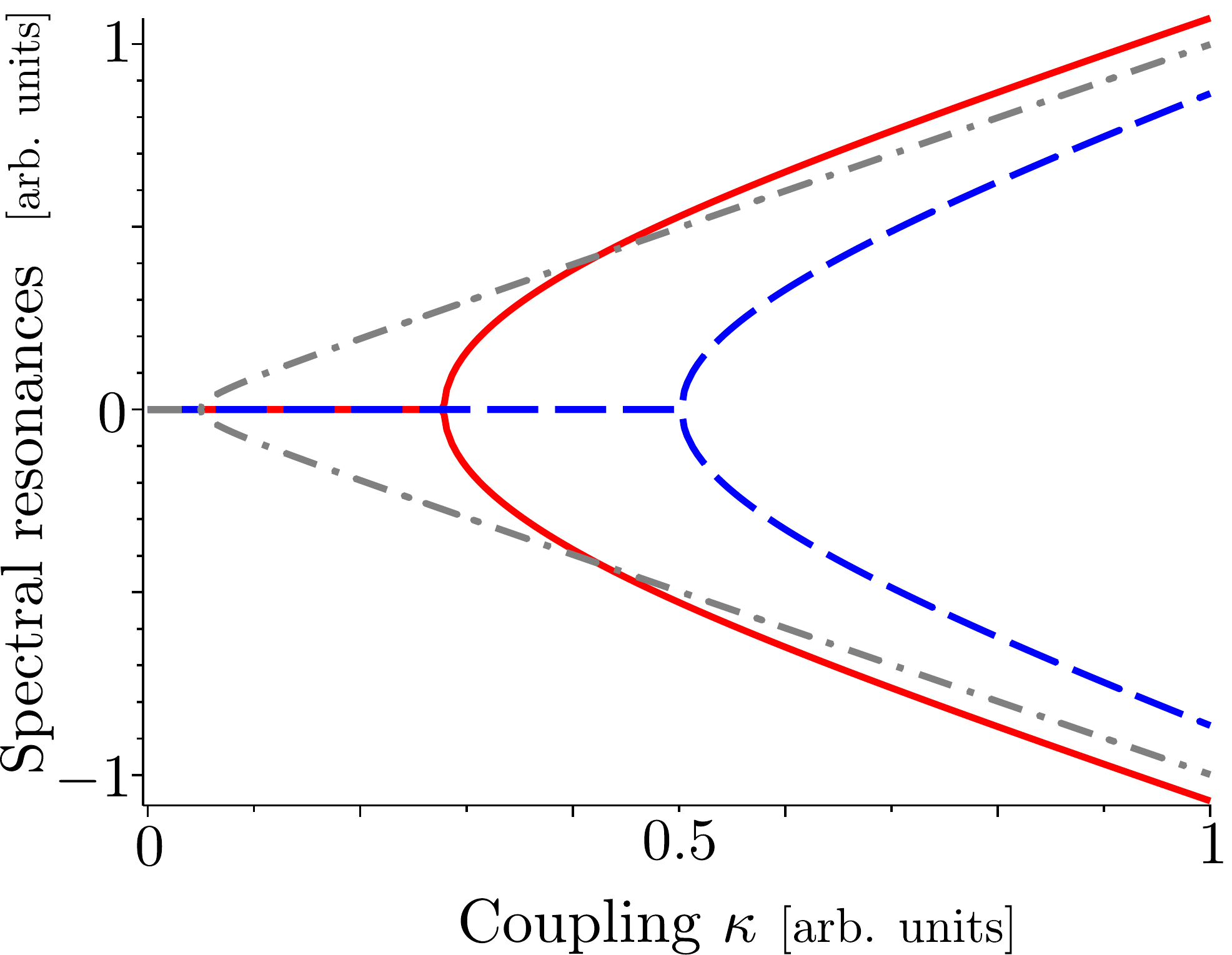}
\caption{The resonances of the power spectra  in the active  (red
solid curve) and  passive (blue dashed curve) cavities from
Fig.~\ref{fig4} (see also Appendix~B, for details). The system
parameters are the same as in Fig.~\ref{fig2}. There is a shift
between mode splittings in the two cavities, which is increasing
with increasing value of waveguide couplings
$\gamma_1=\gamma_2=\gamma$. For comparison, the imaginary
frequencies $\nu_{1,2}$ of the Liouvillian (grey dash-dotted
curves), which are the same as the real frequencies of the NHH
given in Eq.~(\ref{SHeff_eig}), are also displayed on the graph.
The imaginary frequencies of the Liouvillian and resonances of the
emission spectra coincide in the limit $\kappa\to\infty$. On the
other hand, the LEP  of  $\cal L$ in Eq.~(\ref{sLEP1}) and SBPs of
$S_{1,2}$ in Eq.~(\ref{LEP_kappa}) tend to coincide in the limit
$A-\Gamma_1-\Gamma_2\to0$, i.e., in the limit where the assumption
of the linearity of the system can fail.} \label{fig5}
\end{figure}

Nevertheless, the analysis of the power spectra can give us some
additional and valuable  hints to understand the physics of the
system. In Fig.~\ref{fig4} we plot the power spectra of both
cavities for different values of the intercavity coupling
$\kappa$. In Fig.~\ref{fig5}, instead, we plot the peaks of the
power spectra resonances (whose splitting signaling the SBPs) and
the imaginary part of $\nu_{1\, 2}$ associated to the decay of the
TTCF (whose bifurcation indicates the LEP). We chose balanced
intrinsic gain and losses  $A-C_1-C_2=0$ (which is the effective
$\cal PT$-symmetric regime). Thanks to the additional coupling of
the cavities to the waveguides, the total gain in the system
becomes smaller than the total loss i.e.,
$A-\Gamma_1-\Gamma_2=A-C_1-C_2-2\gamma<0$. Our formalism remains
valid for $\gamma$ large enough to ensure that the active cavity
is far below the lasing threshold. As one can see, for very small
values of $\kappa$, the power spectrum in both cavities is
asymmetric, i.e., the emission is mainly observed in the active
cavity, which has a Lorentzian shape [see Fig.~\ref{fig4}(a)].
This is because the coupling is too small for the generated
photons in the active resonator to pass into the passive cavity
and be emitted.
 Again, this is a demonstration of the
impossibility to realize $\cal PT$-symmetry in photonic systems
due to a spontaneous emission enhanced by the gain $A$. If one
were to drive the system by intense classical fields, this would
eventually restore the symmetry, but completely conceal the
presence of the spontaneous-emission fields. Note that similar
conclusions, regarding the self-sustained radiation in the system
and observed asymmetry in the emission spectra, have been
previously obtained in Ref.~\cite{Schomerus2010,Yoo2011}  by applying
scattering theory.

By increasing the coupling strength $\kappa$, the emission
spectrum in the active cavity start exhibiting a
squared-Lorentzian lineshape [see Fig.~\ref{fig4}(b)], which
signals the arising mode splitting in the active resonator, i.e.,
the appearance of an SBP in the system (see Fig.~\ref{fig5}). At
the same time, the emission spectrum in the passive cavity becomes
comparable in power to the power spectrum in the active resonator
but with a Lorentzian lineshape [see Figs.~\ref{fig4}(b) and
\ref{fig5}]. Further increasing  $\kappa$ leads to a clear mode
splitting in the active resonator and the emergence of a
squared-Lorentzian line in the passive resonator [see
Fig.~\ref{fig4}(c) and \ref{fig5}]. For even larger values of
$\kappa$,  $S_1$ and $S_2$ are Lorentzian and coincide with each
other,  showing two well-separated lines, which, in the limit
$\kappa\rightarrow\infty$,  become proportional to the intercavity
splitting $\kappa$  [see Fig.~\ref{fig4}(d) and \ref{fig5}].

\subsubsection{Discussion about the semiclassical limit}
In summary of this section, we have defined and compared the HEP
and one of the LEPs in the semiclassical regime. Whereas the HEP
has been directly obtained from the spectra of the NHH, the LEP
has been determined from the TTCF, which enables to detect well
the LEPs in the system. The analysis provided implies that, in
this regime, both HEP and, at least, one of the LEPs appears for
the same combination of system parameters and has the same decay
rate.  We note that although, in general, one fails to identify
the exact value of an LEP from the power spectra based on the
resonant peaks splitting, it might be possible to detect it by
utilizing other statistical measures applied to the spectra
curves, e.g., such as bimodal coefficients or Binder cumulants.
This study, however, is beyond the scope of this work. Finally, in
the special cases when the system approaches the genuine $\cal
PT$-symmetry with balanced total gain and losses in both cavities,
the mode splitting phenomenon in the power spectra tend to occur
at the exact value of the LEP.

\section{Hamiltonian and Liouvillian exceptional points in the quantum single-photon limit}

Let us consider a situation when there is no more than one photon
in each cavity, i.e., $\langle\hat n_i\rangle\ll1$, $i=1,2$. This
can be easily achieved when the ratio between the gain and the
losses in the active cavity is very low, i.e., $A/\Gamma_1\ll1$,
according to Eq.~(\ref{NSS}).  In this case, the Hilbert space of
the system can be reduced to a four dimensional space, spanned by
the vectors $|j\rangle|k\rangle$ with $j,k=0,1$. As a result, we
can easily represent both  NHH and Liouvillian as small
matrices, allowing their  diagonalization and the study their
EPs in the quantum single-photon limit.

\subsection{Non-Hermitian Hamiltonian exceptional points}

In the two-photon cutoff Hilbert space,  the effective NHH in
Eq.~(\ref{Heff}) attains the following matrix form (see Appendix~C
for details)
\begin{equation}\label{qHeff}  
\hat H_{\rm eff}\equiv\begin{pmatrix}
0 &  & & \\
 & \omega_c-i\frac{\Gamma_2}{2} & i\kappa & \\
& -i\kappa & \omega_c+\frac{i}{2}\left(A-\Gamma_1\right) & \\
& & &2\omega_c+\frac{i}{2}\left(A-\Gamma_{+}\right)
\end{pmatrix},
\end{equation}
with  eigenvalues:
\begin{eqnarray}\label{eta}   
&\eta_{0} = 0, \quad
\eta_{1}=2\omega_c+\frac{i}{2}\left(A-\Gamma_{+}\right), \quad
\eta_{2,3}=\nu_{1,2},
\end{eqnarray}
where $\nu_{1,2}$ are given in Eq.~(\ref{SHeff_eig}).

Note, that because of the resized NHH $\Hef$, compared to that in
Eq.~(\ref{Hef_M}),  apart from the same eigenvalues $\eta_{2,3}$,
this NHH has  also two additional eigenvalues $\eta_0$ and
$\eta_1$.

{Again, because the eigenvalues $\eta_i$, in Eq.~(\ref{eta}), are complex, the NHH $\Hef$ can attain both right and left eigenvectors.}

{The right eigenvectors of the NHH $\Hef$, in Eq.~(\ref{qHeff}), away from HEP, are}
\begin{eqnarray}\label{Heig}  
|\psi_0\rangle&=&|00\rangle, \quad |\psi_1\rangle=|11\rangle,  \nonumber \\
 |\psi_{2,3}\rangle&\equiv&(A-\Gamma_-\pm\beta)|10\rangle+4\kappa|01\rangle,
\end{eqnarray}
where $\beta$ is given below Eq.~(\ref{SHeff_eig}). {The normalization coefficients
for the eigenstates $|\psi_{2,3}\rangle$,  in Eq.~(\ref{Heig}), can be safely dropped, since the considered system in this quantum regime does not
exhibit the \PT-symmetry, where the eigenstates might not be normalized~\cite{Hashimoto2015}.} 

{By inspecting Eq.~(\ref{Heig}), one can clearly see the
coalescence of the eigenvalues $\eta_2=\eta_3=i(A-\Gamma_-)/4$, and
that the eigenvectors $|\psi_{2}\rangle$ and
$|\psi_{3}\rangle$ coalesce to the maximally entangled state $|\psi_{\rm
HEP}\rangle\equiv|10\rangle+|01\rangle$, which occurs at the following
HEP:}
\begin{equation}\label{qHEP}  
 \kappa_{\rm HEP}^{\rm q}=\frac{1}{4}\left|A-\Gamma_-\right|.
\end{equation}
As expected, for this NHH $\Hef$, the HEPs coincide in the
semiclassical and single-photon limits.

{At the HEP, the NHH $\Hef$ becomes non-diagonazible, i.e., it attains a Jordan form. Hence, the generalized eigenspace of the NHH $\Hef$ consists of the the eigenvectors 
\begin{equation} 
|\psi_0\rangle=|00\rangle, \quad |\psi_1\rangle=|11\rangle, \quad |\psi_{\rm HEP}\rangle\equiv|10\rangle+|01\rangle,
\end{equation}
and the singlet-type pseudo-eigenvector~\cite{Hashimoto2015}:
\begin{equation} 
|\psi'_{\rm HEP}\rangle\equiv |10\rangle-|01\rangle.
\end{equation}
}

\subsection{Liouvillian exceptional points}
\subsubsection{Eigenvalues}
Within the two-photon approximation, the Liouvillian $\cal L$ in
Eq.~(\ref{MESR}) is a $4\times4$ matrix. By combining together
Eqs.~(\ref{MESR}) and~(\ref{LS}), one obtains the following
eigenvalues of $\cal L$ (see Appendix~C for details):
\begin{eqnarray}\label{lambda}  
&\lambda_0=0, \quad \lambda_{1,2}=i\omega_c-\frac{1}{2}A_{+}+\frac{1}{4}E_{\pm},& \nonumber \\
&\lambda_{3,4}=-\frac{1}{2}(A_{+}\pm D), \quad \lambda_{5,6}=-\frac{1}{2}A_{+},&\nonumber \\
&\lambda_7=2i\omega_c-\frac{1}{2}A_{+}, \quad \lambda_{8,9}=i\omega_c-\frac{1}{2}A_{+}-\frac{1}{4}E_{\pm},&\nonumber \\
& \lambda_{10}=-A_{+}, \quad \lambda_{11,12,13,14,15}=\lambda_{1,2,7,8,9}^*,&
\end{eqnarray}
where  $D=\sqrt{A_{-}^2-16\kappa^2}$, $A_{\pm}=A+\Gamma_{\pm}$,
\begin{equation*}
E_{\pm}=\sqrt{2}\sqrt{(A+\Gamma_1)^2+\Gamma_2^2-16\kappa^2\pm F},
\end{equation*}
 and
\begin{equation}
F=\Big(A_{+}^2A_{-}^2+16\kappa^2\left(8A\Gamma_2-A_{+}^2\right)\Big)^{\frac{1}{2}}.
\nonumber
\end{equation}

{As an example, we plot the frequency spectrum $\lambda_i$ of the Liouvillian in Fig.~\ref{fig6}.}

{Note that the Liouvillian frequency spectrum, in general, strongly depends on the interaction $\kappa$ between the fields in the two cavities, particularly, when $\Gamma_i\gg\Gamma_j$, $i,j=1,2$, $i\neq j$.
It means that compared to the case when both cavities are isolated from each other, the decay rates $\lambda_i$ of the Liouvillian states can either be substantially facilitated or impeded by this interaction~\cite{Cai2013,Poletti2013,Bouganne2019}. 
}

\subsubsection{Eigenmatrices}
The eigenmatrices $\hat\rho_i$, corresponding to the real-valued
eigenvalues $\lambda_i$, can be written as follows:
\begin{equation}\label{r01}  
\hat\rho_j=\frac{1}{N_j}\begin{pmatrix}
\rho^{(j)}_{00} & & & \\
& \rho^{(j)}_{01} & & \\
& &  \rho^{(j)}_{10} & \\
& & &  \rho^{(j)}_{11}
\end{pmatrix}, \quad j=0,10,
\end{equation}
\begin{eqnarray}\label{r23}  
\hat\rho_{3,4}=\begin{pmatrix}
\rho'_{00}\pm f_1D & & & \\
& \rho'_{01}\pm f_2D  & \rho'_{0110}\pm{f_3}{D}& \\
&\rho'_{0110}\pm{f_3}{D} &  \rho'_{10}\pm f_4D  & \\
& & &  \rho_{11}
\end{pmatrix}, \nonumber \\
\end{eqnarray}
\begin{eqnarray}\label{r56}  
\hat\rho_{5}=\begin{pmatrix}
\rho_{00}& & & \\
& \rho_{01} & \rho_{0110}& \\
&\rho_{0110} &  \rho_{10}  & \\
& & &  \rho_{11}
\end{pmatrix}, \quad \hat\rho_{6}=\begin{pmatrix}
0 & &  \\
&\hat\sigma_{y} &  \\
&& 0\\
\end{pmatrix},   \nonumber \\
\end{eqnarray}
where $D$ is given in Eq.~(\ref{lambda}), $\hat\sigma_{y}$ is
$2\times2$ Pauli matrix, and the rest parameters are given in
Appendix~C.

The remaining non-Hermitian eigenmatrices with complex eigenvalues
are the following:
\begin{eqnarray}\label{r4567}  
&\hat\rho_{k}=\begin{pmatrix}
0 & \rho_{0001}(\lambda_k)  &  \rho_{0010}(\lambda_k) & \\
&  0& 0 &\rho_{0111} \\
& &  0 &\rho_{1011}(\lambda_k) \\
& & &  0
\end{pmatrix},& \nonumber \\
 &k=1,2,8,9,  \quad \text{and} \quad
 \hat\rho_{7}=\begin{pmatrix}
 &   &  & 1\\
&  & & \\
& &  & \\
& & &
\end{pmatrix}, &
\end{eqnarray}
and for the eigenvalues $\lambda_l$ with $l=11,12,13,14,15$, the
eigenmatrices are found as a Hermitian conjugate of the
eigenmatrices $\hat\rho_{k}$, with $k=1,2,7,8,9$, respectively,
where $\hat\rho_{k}$ are given in Eq.~(\ref{r4567}). The exact
values of all the eigenmatrices in Eqs.~(\ref{r01})--(\ref{r4567})
are given in Appendix~C. Obviously, the spectrum of the
Liouvillian $\cal L$ is much richer than that of the NHH $\hat
H_{\rm eff}$.

\subsubsection{Spectral decomposition and LEPs}
{{\it (1) Study of $\hat\rho_{0,10}$}.---
The Hermitian diagonal eigenmatrix $\hat\rho_0$, in
Eq.~(\ref{r01}), is the steady-state density matrix. As expected,
the steady state is nothing else but a classical mixture of the
states $|jk\rangle\langle jk|$, where $j,k=0,1$. The Hermitian
eigenmatrix $\hat\rho_{10}$, instead, is responsible for the
dynamical evolution of the diagonal elements  $|jk\rangle\langle jk|$
towards the steady state with the decaying rate $\lambda_{10}$.}

 {\it (2) Study of $\hat\rho_{3,4}$}.--- Let us now study the eigenmatrices
$\hat\rho_{3,4}$, since, as it will be shown below, their eigenstates are the closest to those defined in
Eq.~(\ref{Heig}). 

{As it was stressed earlier, the EP of the Liouvillian is defined as a point in the parameter space where the eigenvalues
and eigenmatrices of $\mathcal{L}$ coincide. By inspection of Eqs.~(\ref{lambda}) and (\ref{r23}), one can see that both eigenvalues $\lambda_{3,4}$ and corresponding eigenmatrices $\hat\rho_{3,4}$ coincide whenever $D=0$. 
Moreover, the eigenvalues $\lambda_{3,4}$ and eigenmatrices $\hat\rho_{3,4}$ coalesce with the eigenvalue $\lambda_5$, in Eq.~(\ref{lambda}), and the eigenmatrix $\hat\rho_5$, in Eq.~(\ref{r56}), respectively.
Therefore, the Liouvillian $\cal L$ acquires a third-order EP given by
\begin{equation}\label{qLEP1}  
\kappa_{{\rm LEP},1}^{\rm
q}=\frac{1}{4}|A_{-}|=\frac{1}{4}\left|A+\Gamma_1-\Gamma_2\right|.
\end{equation}
The subscript $1$ at $\kappa_{{\rm LEP},1}^{\rm
q}$ stands for the first LEP, since as it will be shown below, there are at least  two LEPs in the system,  and which in the limit $A\to0$ coincide.
}

\begin{figure}[tb] 
\includegraphics[width=0.45\textwidth]{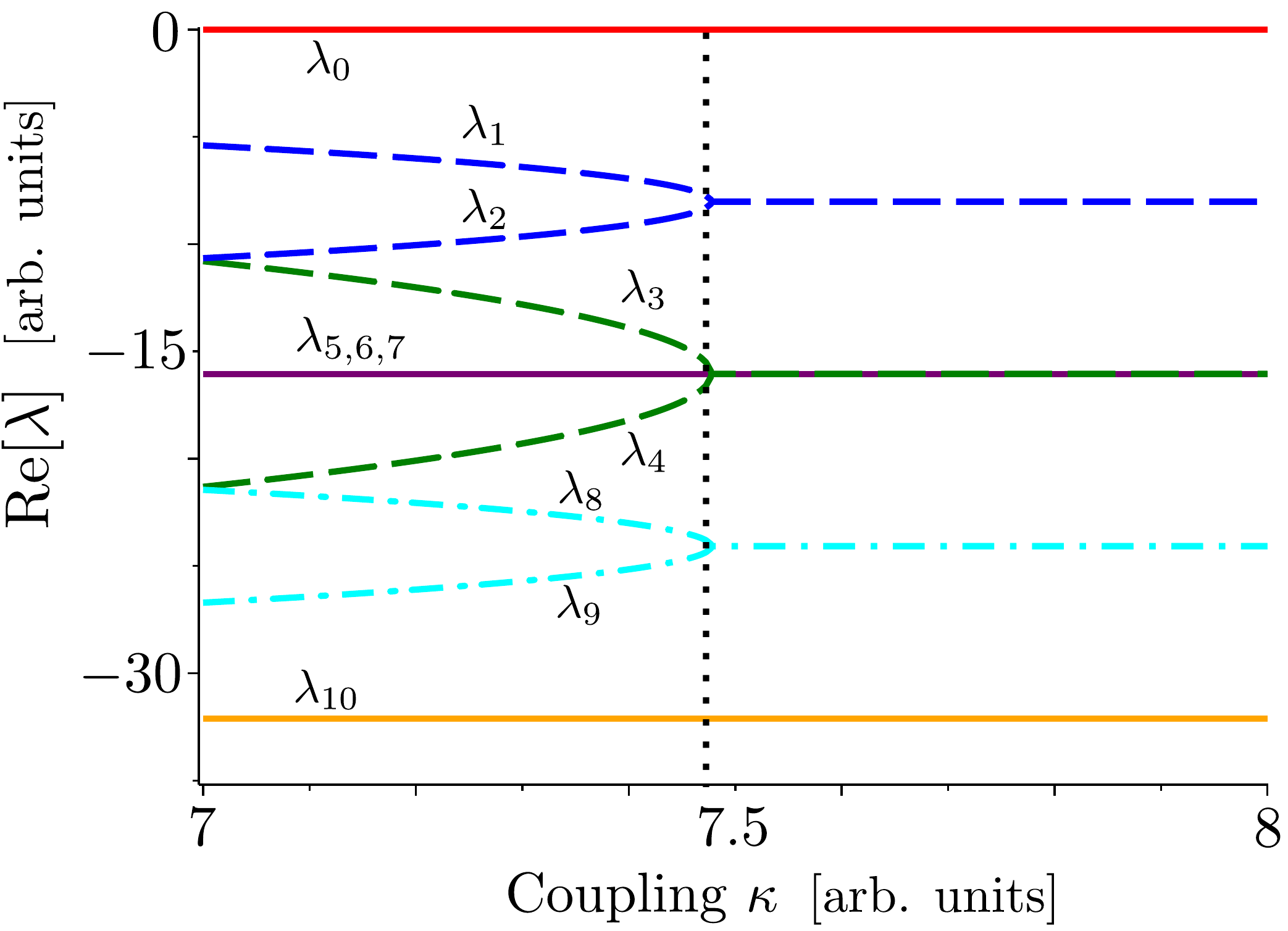}
\caption{ Liouvillian EPs and the real part of its eigenvalues
$\lambda_j$, according to Eq.~(\ref{lambda}): ${\rm
Re}[\lambda_{0}]$ (red solid curve), ${\rm Re}[\lambda_{1,2}]$
(blue dash-dotted curves), ${\rm Re}[\lambda_{3,4}]$ (green dashed
curves),  ${\rm Re}[\lambda_{5,6,7}]$ (purple solid curve), ${\rm
Re}[\lambda_{8,9}]$ (cyan dotted curves), and ${\rm
Re}[\lambda_{10}]$ (orange solid curve). The gain in the active
cavity $A=0.01$ [arb.~units], while the losses in the active and passive
cavities are the same as in Fig.~\ref{fig2}. The maximum value of
the mean photon number in the active cavity is $\langle\hat
n_1\rangle_{\rm max}\approx3\cdot10^{-4}$. For comparison, the HEP
(vertical grey dotted line)  of the NHH, given in
Eq.~(\ref{qHEP}), is also displayed. This graph indicates
that, in the single-photon regime, the LEPs and HEPs tend to
coincide. Moreover, as it follows from the plot,  the values of
the two LEPs, given in Eqs.~(\ref{qLEP1}) and~(\ref{qLEP_2}), also
show the tendency to overlap. } \label{fig6}
\end{figure}

\begin{figure}[tb] 
\includegraphics[width=0.48\textwidth]{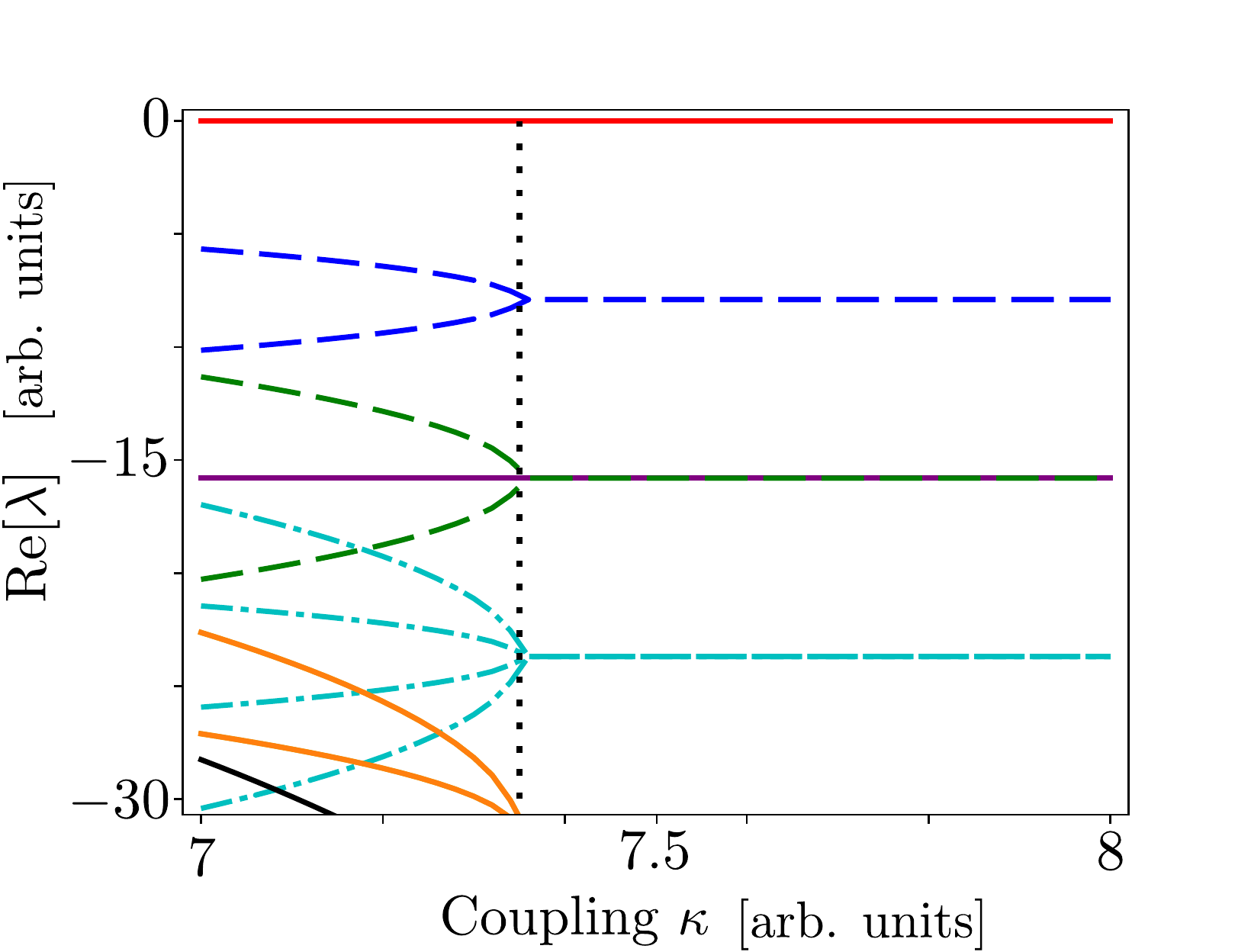}
\caption{ Liouvillian EPs and the real part of its eigenvalues
$\lambda$ for a multiphoton system with up to eight photons in
each cavity.  The system parameters are: the gain in the active
cavity $A=0.5$ [arb.~units], the losses in the active and passive
cavities are the same as in Fig.~\ref{fig2}. For comparison, the
HEP of the NHH (vertical grey dotted line), given in
Eq.~(\ref{qHEP}), is also displayed. This graph indicates
that with an increasing photon number in the system,  the LEPs and
HEP tend to coincide as in Fig.~\ref{fig6}.} \label{fig7}
\end{figure}

{Remarkably, despite the fact that the HEP $ \kappa_{\rm HEP}^{\rm q}$ and  LEP $\kappa_{{\rm LEP},1}^{\rm
q}$ are of different order and have a slightly different form (opposite signs at $\Gamma_1$ and $\Gamma_2$), they occur for the same combination of parameters in the weak-gain regime, where the two-photon cutoff can be safely applied.}
Namely, when considering a two-photon cutoff, one must bear in mind that the
gain $A$ in the active cavity should be very small compared to the
total losses in the active cavity, i.e., $A/\Gamma_1\ll1$, in
order to justify the two-photon approximation. Therefore, in the
case, when $A$ becomes negligible compared to both $\Gamma_1$ and
$\Gamma_2$, the LEP and HEP tend to coincide, i.e., $\kappa_{\rm
HEP}^{\rm q}\cong\kappa_{{\rm LEP},1}^{\rm q}$  (see also
Fig.~\ref{fig6}). Most importantly, our numerical results also
indicate that even by increasing the gain $A$, and enlarging the
subspace of the Hilbert space to higher-photon excitations, the
LEP and HEP demonstrate the same tendency to overlap, i.e.,
$\kappa_{{\rm LEP},1}^{\rm q}\to\kappa_{{\rm HEP}}^{\rm q}$ with
increasing $\langle\hat n_1\rangle$ (see also Fig.~\ref{fig7}).
Therefore, the same EP can have {\it different} order for the NHH $\Hef$ and the Liouvillian $\cal L$.

{Note that the previous discussion can also be generalized if we consider a \emph{truly} two coupled two-level system.
Namely, if instead of considering the small-gain regime of a bosonic system we take under consideration a system where
the photon-photon interaction determines a photon-blockade regime, the NHH not only fails to capture the nature of the LEP, 
but also the parameters for which it occurs.}

When $\kappa<\kappa_{{\rm LEP},1}^{\rm q}$,
both eigenmatrices $\hat\rho_{3,4}$ are Hermitian, and one can
immediately find their eigenstates as follows
\begin{eqnarray}\label{eig_r23}  
|\psi^{(3,4)}_0\rangle&=&|00\rangle, \quad |\psi^{(3,4)}_1\rangle=|11\rangle, \nonumber\\
|\psi^{(3)}_{2,3}\rangle&\equiv&{4\kappa}|01\rangle+\Big({D\pm|A_-|}\Big)|10\rangle, \nonumber \\
|\psi^{(4)}_{2,3}\rangle&\equiv&{4\kappa}|01\rangle+\Big({-D\pm|A_-|}\Big)|10\rangle.
\end{eqnarray}
Direct inspection of Eq.~(\ref{eig_r23}) reveals that  the
subspace of the eigenstates of the density matrices
$\hat\rho^{(2,3)}$ resembles the space of the eigenstates of the
NHH $\hat H_{\rm eff}$ in Eq.~(\ref{Heig}). Moreover, in the limit $A\to0$, this resemblence turns into equivalence. 

{When $\kappa=\kappa_{{\rm LEP},1}^{\rm q}$,  then $\lambda_{3}=\lambda_4=\lambda_5=\lambda_{\rm EP}=-{A_+}/{2}$, and $\hat\rho_{3}=\hat\rho_4=\hat\rho_5$ (see also Fig.~\ref{fig6}).  The latter implies that the eigenstates of the Liouvillian at this LEP,  which belong to the eigenmatrix $\hat\rho_5$ and describe the intercavity fields interaction,  are the maximally entangled states,  according to  Eq.~(\ref{psi6}).
Additionally, at the LEP $\kappa_{{\rm LEP},1}^{\rm q}$, the algebraic multiplicity of the eigenvalue $\lambda_{\rm LEP}$ exceeds its geometric multiplicity, according to Eqs.~(\ref{lambda}), (\ref{r23}) and (\ref{r56}). 
Namely, the algebraic multiplicity of  $\lambda_{\rm LEP}$ becomes four, but geometric multiplicity equals two,  because there are only two linearly independent eigenmatrices $\hat\rho_{5,6}$ for this eigenvalue. The rank of the eigenmatrices $\hat\rho_{3,4,5}$ is the same and equals four, whereas the rank of the eigenmatrix $\hat\rho_6$ equals two. 
Therefore, one has to find two additional generilized pseudo-eigenmatrices of the rank four for the  Liouvillian $\cal L$, which takes on a Jordan form in this case.  These pseudo-eigenmatrices, denoted as  $\hat\rho'_5$ and $\hat\rho''_5$, can be found via Jordan chain relations (see also Appendix D, for details).  When found,  the density matrix $\hat\rho(t)$ of the system can be decomposed in the form given in Eq.~(\ref{rho_diag_EP}), with an additional contribution $c''_{\rm EP}(t)\hat\rho''_{\rm EP}$, where $\hat\rho''_{\rm EP}=\hat\rho''_5$ and  $c''_{\rm EP}(t)=t^2\exp(\lambda_{\rm EP}t){\rm Tr}[\hat\sigma''_5\hat\rho(0)]$. 
}

When $\kappa>\kappa_{{\rm LEP},1}^{\rm q}$, one has to consider
the symmetric $\hat\rho^{\rm s}_{3,4}$ and antisymmetric
$\hat\rho^{\rm a}_{3,4}$  density matrices, as was explained
above. Thus, one eventually finds the form of the eigenstates for
the symmetric density matrices $\hat\rho^{\rm s}_{3,4}$:
\begin{eqnarray}\label{sym}  
|\psi^{(3,4)}_0\rangle_{\rm s}&=&|00\rangle, \quad |\psi^{(3,4)}_3\rangle_{\rm s}=|11\rangle, \nonumber\\
|\psi^{(3,4)}_{1,2}\rangle_{\rm
s}&\equiv&{-\delta}|01\rangle+\Big({D^2\pm\sqrt{\delta^2+D^4}}\Big)|10\rangle,
\end{eqnarray}
where $\delta=4\kappa A_-$. The antisymmetric matrices
$\hat\rho^{\rm a}_{3,4}$, instead, have the following eigenstates
\begin{eqnarray}\label{anti}  
|\psi^{(3,4)}_0\rangle_{\rm a}&=&|00\rangle, \nonumber \\
|\psi^{(3,4)}_{1,2}\rangle_{\rm
a}&\equiv&{-4\kappa}|01\rangle+\Big({A_-\pm\gamma}\Big)|10\rangle,
\end{eqnarray}
where $\gamma=\sqrt{16\kappa^2+A_-^2}$. As one can see from
Eqs.~(\ref{sym}) and~(\ref{anti}), the eigenstates of
$\hat\rho^{\rm s,a}_{3}$ and $\hat\rho^{\rm s,a}_{4}$ are the
same. This stems from the fact that
$\hat\rho_4=\hat\rho_3^{\dagger}$ according to Eq.~(\ref{r23}), in
the case when $\kappa>\kappa_{{\rm LEP},1}^{\rm q}$.  As both
Eqs.~(\ref{sym}) and~(\ref{anti}) infer, in this case, there is also no
exact matching between the eigenstates of $\hat H_{\rm eff}$ and
$\hat\rho^{\rm s,a}_{3,4}$ of the Liouvillian $\cal L$;  thus,
providing a different description of the interaction between the
cavities. In the limit $\kappa\to\infty$,  the two antisymmetric
intercavity eigenstates reduce to $|\psi^{(3,4)}_{1,2}\rangle_{\rm
a}\equiv|10\rangle\pm|01\rangle$, whereas the symmetric
intercavity eigenstates $|\psi^{(3,4)}_{2,3}\rangle$ reduce  to either $|01\rangle$ or
$|10\rangle$. According to Eqs.~(\ref{rho_diag}) and ~(\ref{lambda}), away from the
EPs,  the elements $|\psi_n^{(3,4)}\rangle_{\rm s,a}\langle\psi_n^{(3,4)}|$ of
the eigenmatrices $\hat\rho^{\rm s,a}_{3,4}$ in Eqs.~(\ref{sym})
and (\ref{anti}), apart from the exponential decay, also acquire an oscillating term proportional to $D$.

 {\it (3) Study of $\hat\rho_{1,2,8,9}$}.--- Now let us focus
on the non-Hermitian eigenmatrices $\hat\rho_i$, $i=1,2,8,9$,
given in Eq.~(\ref{r4567}). These eigenmatrices define the second
LEP in the system:
\begin{eqnarray}\label{qLEP_2} 
\kappa_{{\rm LEP},2}^{\rm
q}&=&\frac{\left|(A+\Gamma_1)^2-\Gamma_2^2\right|}{4\sqrt{A_{+}^2-8A\Gamma_2}}.
\end{eqnarray}
At  the LEP $\kappa_{{\rm LEP},2}$, one  can observe the
coalescence of the eigenmatrices $\hat\rho_{1}$ and
$\hat\rho_{2}$, as well as the coalescence of the eigenmatrices
$\hat\rho_{8}$ and $\hat\rho_{9}$, and the same applies to their
Hermitian conjugate (see Fig.~\ref{fig6}). Thus, the LEP $\kappa_{{\rm LEP},2}^{\rm
q}$ is of the {\it second order}. In particular, when $A\ll\Gamma_{1,2}$, which
is true in the two-photon cutoff, the LEP $\kappa_{{\rm
LEP},2}^{\rm q}$ is also inclined to coincide with  $\kappa_{{\rm
LEP},1}^{\rm q}$ and $\kappa_{\rm HEP}^{\rm q}$ (see
Fig.~\ref{fig6}). Importantly, the same conclusion, regarding the
convergence of the LEPs to the HEP, remains valid even when we try
to increase the gain $A$, i.e., by extending the Hilbert space to
larger photon numbers (see Fig.~\ref{fig7}).

By performing the eigen-decomposition of  the Hermitian symmetric
and antisymmetric eigenmatrices $\hat\rho_i^{\rm s,a}$,
$i=1,2,8,9$, the corresponding wave functions
$|\psi_n^{(i)}\rangle$, in general, take the form of the following
superpositions $|\psi_n^{(\rm s,a)}\rangle=\sum
c_{ij}|i\rangle|j\rangle$.  Moreover, away from the EPs, the
eigenmatrices elements  $|\psi_n^{(\rm
s,a)}\rangle\langle\psi_n^{(\rm s,a)}|$, in addition to the
gradual decay, rapidly oscillate around the cavity resonance
frequency $\omega_c$, according to Eqs.~(\ref{rho_diag})
and~(\ref{lambda}).

{{\it (4) Study of $\hat\rho_{5,6}$.---} The real eigenvalue
$\lambda_{5,6}$ in Eq.~(\ref{lambda}) has both algebraic and
geometric multiplicity of two. This means that there are two
linearly independent eigenmatrices corresponding to this
eigenvalue, and which are given in Eq.~(\ref{r56}). 
 The Hermitian
non-diagonal eigenmatrix $\hat\rho_5$, along with the eigenstates $|00\rangle$ and $|11\rangle$,
has the following  intercavity  maximally entangled states:
\begin{equation}\label{psi5} 
 |\psi^{(5)}_{1,2}\rangle\equiv |01\rangle\pm |10\rangle.
\end{equation}
On the other hand, the eigenmatrix $\hat\rho_6$ possesses only the following 
entangled states:
\begin{equation} \label{psi6} 
 |\psi^{(6)}_{1,2}\rangle\equiv|01\rangle\pm i|10\rangle.
\end{equation}
The elements $|\psi^{(j)}_n\rangle\langle\psi^{(j)}_n|$ of the eigenmatrix $\hat\rho_j$, $j=5,6$, decay in time with the rate $\lambda_{5,6}$.
}

{\it (5) Study of $\hat\rho_{7}$.---} Finally, we find that the
non-Hermitian eigenmatrices $\hat\rho_7$ and
$\hat\rho_7^{\dagger}$ give the following  intercavity eigenstates
\begin{equation}\label{psi7}  
|\psi^{(7)}_{1,2}\rangle\equiv|00\rangle\pm|11\rangle.
\end{equation}
The products $|\psi^{(7)}_{1,2}\rangle\langle\psi^{(7)}_{1,2}|$,
which constitute the eigenmatrix $\hat\rho_7$,   also decay with the same rate as the states $\hat\rho_{5,6}$, but
oscillate at the double frequency $2\omega_c$, according to
Eqs.~(\ref{rho_diag}) and~(\ref{lambda}).

 {\it (6) General discussion about the spectral
decomposition}.--- {In the single-photon limit, the LEPs and HEPs
tend to coincide, as in the semiclassical case for many photons.
On the other hand, the spectral properties of the Liouvillian
drastically differ from those of the NHH and exhibit a rich
dynamical nature.{ Most importantly, even if the LEPs and HEPs coincide for the 
same set of the system parameters, they can have completely different order, thus, pointing to the different nature of HEPs and LEPs.
}

\section{Conclusions}

We have studied the quantum and semiclassical exceptional points
of a linear non-Hermitian system of coupled cavities with losses
and gain within the Scully-Lamb quantum laser model. Specifically,
we have found the expressions for the HEPs and LEPs of the
non-Hermitian system  in both semiclassical and quantum regimes,
i.e., when the system contains either classical fields with many
photons or single photons, respectively. Our results have
demonstrated that in either regime {the position of} both HEPs and
LEPs tend to be the same. {Moreover, physical quantities such as
the decay rates of the first order correlation functions are the
same.} In the semiclassical regime, we have calculated the HEP
from the spectra of the effective non-Hermitian Hamiltonian,
whereas the LEP has been determined from the two-time correlation
function. {Importantly, our  analysis has also revealed that
it is exactly a TTCF that enables to identify a true LEP in the
semiclassical regime, whereas the field power spectra, in general,
{\it fail} to reveal the exact value of the LEP}.
  In the quantum mode, we have assumed that the system
contains no more than one photon in each cavity;  thus,
allowing us to write down both the NHH and Liouvillian in a finite
matrix form. Our calculations have also indicated that whereas the parameters for which
HEPs and LEPs can coincide, the spectral structure of the
Liouvillian is much richer compared to the NHH, revealing its full
dynamical nature. {Moreover, we have found that, in the quantum regime, the
very order of EPs can be different for HEPs and LEPs, respectively, with LEPs being in general of higher order.}

\begin{acknowledgments}

The authors kindly acknowledge Alberto Biella, Nicola Bartolo,
{\c{S}}ahin K. \"{O}zdemir, and Jan Pe\v{r}ina Jr. for insightful
discussions. I.A. thanks the Grant Agency of the Czech Republic
(Project No.~17-23005Y), the Project
CZ.02.1.01\/0.0\/0.0\/16\_019\/0000754. F.M. is supported by the
FY2018 JSPS Postdoctoral Fellowship for Research in Japan. F.N. is
supported in part by the: MURI Center for Dynamic Magneto-Optics
via the Air Force Office of Scientific Research (AFOSR)
(FA9550-14-1-0040), Army Research Office (ARO) (Grant No.
W911NF-18-1-0358), Asian Office of Aerospace Research and
Development (AOARD) (Grant No. FA2386-18-1-4045), Japan Science
and Technology Agency (JST) (via the Q-LEAP program, and the CREST
Grant No. JPMJCR1676), Japan Society for the Promotion of Science
(JSPS) (JSPS-RFBR Grant No. 17-52-50023, and JSPS-FWO Grant No.
VS.059.18N), and the RIKEN-AIST Challenge Research Fund.

\end{acknowledgments}

\section*{APPENDICES}

\appendix

\section{Some remarks regarding the use of quantum Langevin forces in Sec.~IIIA}

Here, we would  like to make a few comments regarding the
widespread use of quantum Langevin forces, given in
Eq.~(\ref{LE}), and which encompass the quantum noise in the
system.

In the usual approach, applied in the related
literature~\cite{Kepesidis2016,Ghamsari2017}, especially devoted
to the $\cal PT$-symmetric cavities, one may encounter the
following Langevin equations for the quantum fields $\hat a_1$ and
$\hat a_2$ in the coupled cavities  (ignoring the complex
frequency part):
\begin{eqnarray}\label{A1} 
\frac{{\rm d}}{{\rm d}t}\hat a_1&=&\frac{g_1}{2}\hat a_1-\kappa\hat a_2+\sqrt{g_1}\hat f_1^{\dagger}, \nonumber \\
\frac{{\rm d}}{{\rm d}t}\hat a_2&=&-\frac{g_2}{2}\hat
a_2+\kappa\hat a_1+\sqrt{g_2}\hat l_2,
\end{eqnarray}
where $g_1>0$ ($g_2>0$) describes amplification (damping) in the
active (passive) cavity, and $\hat f_j^{\dagger}$ ($\hat l_j$) is
the quantum Langevin force describing quantum noise amplification
(dissipation) in the $j$th cavity.  Moreover, one applies the
Markovian approximation, i.e.,
\begin{equation} 
[\hat O_j(t),\hat O_k^{\dagger}(t')]=\delta_{jk}\delta(t-t'),
\end{equation}
where $\hat O=\hat f$, and $\hat l,  j=1,2$.

 In the case when there are no thermal photons in the environment, one obtains
\begin{equation} 
\langle\hat f_j(t)\hat f_j^{\dagger}(t')\rangle=\langle\hat
l_j(t)\hat l_j^{\dagger}(t')\rangle=\delta(t-t').
\end{equation}
For the case when $\kappa=0$, by direct calculation using
Eq.~(\ref{A1}), one acquires the following expression for the mean
photon number in the active cavity:
\begin{equation} 
\langle\hat n_1(t)\rangle=\exp\left(2g_1t\right)-1.
\end{equation}
 Needless to say, the last expression diverges in the limit $t\rightarrow\infty$. In this case, one needs to incorporate a nonlinear term in the first equation in Eq.~(\ref{A1}) accountable for gain saturation.

For the case when the active cavity is below the lasing threshold,
and again assuming $\kappa=0$,  by blindly replacing the gain
$g_1$  in Eq.~(\ref{A1}) by the net negative gain
$g_1=A-\Gamma_1<0$, where $A$ is the total gain, and $\Gamma_1$ is
the total loss in the active cavity, one obtains the unphysical
solution with $\langle\hat n_1(t)\rangle<0$. To resolve the latter
problem, one has to modify Eq.~(\ref{A1}) with an additional noise
operator $\hat l_1$ responsible for dissipation, i.e.,
\begin{eqnarray}\label{LE2} 
\frac{{\rm d}}{{\rm d}t}\hat a_1&=&\frac{A-\Gamma_1}{2}\hat a_1-\kappa\hat a_2+\sqrt{A}\hat f_1^{\dagger}+\sqrt{\Gamma_1}\hat l_1, \nonumber \\
\frac{{\rm d}}{{\rm d}t}\hat a_2&=&-\frac{\Gamma_2}{2}\hat
a_2+\kappa\hat a_1+\sqrt{\Gamma_2}\hat l_2.
\end{eqnarray}
Now, the rate equations in the form given in Eq.~(\ref{LE2})
provide the same spectral properties of the system as the rate
equations derived from the linear Scully-Lamb ME in
Eq.~(\ref{MESR}).

It is important to stress that even the Langevin equations in
Eq.~(\ref{LE}) for the effective NHH $\Hef$, given in
Eq.~(\ref{Heff}), may lead to erroneous results when the laser
cavity operates near the threshold. In this case, it is a
necessity to apply the general Scully-Lamb ME in
Eq.~(\ref{MES})~\cite{ScullyLambBook}.

\section{Some additional calculations provided for Sec.~IIIB}

\subsection{Coefficients for the TTCFs in Eq.~(\ref{TTCFf})}

The coefficients $u_{1,2}$ and $v_{1,2}$ in Eq.~(\ref{TTCFf}) have
the following forms:
\begin{eqnarray} 
u_{1,2}=&&\frac{-A}{2N}\Big[\Big(\Gamma_2(A-\Gamma_+)-4\kappa^2\Big)\beta \nonumber \\
&&\pm(A-\Gamma_+)\Big(\Gamma_2(A-\Gamma_-)-4\kappa^2\Big)\Big], \nonumber \\
v_{1,2}=&&\frac{2A\kappa^2}{N}\Big[\beta\pm(A-\Gamma_+)\Big],
\end{eqnarray}
where
\begin{equation}  
N=(A-\Gamma_+)\Big[(A-\Gamma_1)\Gamma_2-4\kappa^2\Big]\beta.
\end{equation}

\subsection{Formulas for constants $P_i$ and $Q_i$ in Eq.~(\ref{PQ})}

For the TTCFs $\langle\hat a_i^{\dagger}(0)\hat
a_i(\tau)\rangle_{\rm ss}$, $i=1,2$, the expressions for $P_i$ and
$Q_i$ become
\begin{eqnarray} 
P_1&=&-A\frac{4\Gamma_2^2+(A-\Gamma_+)^2}{(A-\Gamma_+)^3}, \nonumber \\
Q_1&=&-A\frac{(A-\Gamma_+)(A-\Gamma_-)(A-\Gamma_1-3\Gamma_2)}{4(A-\Gamma_+)^3}, \nonumber \\
P_2&=&-A\frac{(A-\Gamma_-)^2}{(A-\Gamma_+)^3}, \nonumber \\
Q_2&=&A\frac{(A-\Gamma_-)^2}{4(A-\Gamma_+)^2}.
\end{eqnarray}
For the linear system under consideration, the following condition
$(A-\Gamma_+<0)$ is always satisfied. The latter implies that the
constants $P_{1,2}$ and $Q_2$ are always positive-valued. On the
other hand, the positivity (negativity) of the constant $Q_1$ is
determined by the positivity (negativity) of the expression
$A-\Gamma_-$, which can be either positive or negative.

\subsection{Resonant frequencies of the power spectra $S_1$ and $S_2$ presented in Fig.~\ref{fig5}.}

The frequencies of the resonant peaks in the emission spectra
$S_j(\omega)$ can be found as the maxima of the functions
$S_1(\omega)$ and $S_2(\omega)$. By solving the equations
\begin{equation*}
\frac{{\rm d}S_j(\omega)}{{\rm d}\omega}=0, \quad j=1,2,
\end{equation*}
 with respect to $\omega$ one finds the following relations for the spectral peaks in both cavities:
\begin{eqnarray}\label{B4}  
\omega_{1}^{\pm} &=& \omega_c\pm\frac{1}{2}{\rm Re}\left[\left(2\kappa\sqrt{4\kappa^2+2\Gamma_2(\Gamma_2-G_1)}-\Gamma_2^2\right)^{\frac{1}{2}}\right], \nonumber \\
\omega_{2}^{\pm} &=& \omega_c\pm\frac{1}{4}{\rm
Re}\left[\sqrt{16\kappa^2-2(G_1^2+\Gamma_2^2)}\right]. \label{W2}
\end{eqnarray}

From Eq.~(\ref{B4}), one can easily find the conditions at which
the two resonant peaks coalesce in either cavity, as given in
Eq.~(\ref{LEP_kappa}).

\section{Liouvillian eigenmatrices $\hat\rho_i$ given in Eqs.~(\ref{r01})--(\ref{r4567})}

Within the effective Hilbert space spanned by the vectors
$|jk\rangle$, $j,k=0,1$, the annihilation boson operators for the
fields $\hat a_1$ and $\hat a_2$  in the active and passive
cavities take the following matrix forms
\begin{equation}\label{C1} 
\hat a_1=\begin{pmatrix}
0 & 1 \\
0 & 0
\end{pmatrix}\otimes
\hat I, \quad \hat a_2=\hat I\otimes\begin{pmatrix}
0 & 1 \\
0 & 0
\end{pmatrix},
\end{equation}
respectively, where $\hat I$ is the $2\times2$ identity matrix. By
using the matrix representation of the boson operators  in
Eq.~(\ref{C1}), one can straightforwardly calculate the
eigenvalues and eigenmatrices of the Liouvillian $\cal L$ in
Eqs.~(\ref{MESR}) and~(\ref{LS}). Below, we write the elements of
the Liouvillian eigenmatrices $\hat\rho_j$ given in
Eqs.~(\ref{r01})--(\ref{r4567}).

\subsection{Liouvillian eigenmatrix $\hat\rho_0$ in Eq.~(\ref{r01})}

The elements of the steady-state eigenmatrix $\hat\rho_{0}$, given
in Eq.~(\ref{r01}), are
\begin{eqnarray} 
\rho_{00}&=&\Gamma_1\Gamma_2A_{+}^2+4\kappa^2\Gamma_{+}^2, \nonumber \\
\rho_{01}&=&4A\kappa^2\Gamma_{+}, \nonumber \\
\rho_{10}&=&A\left(\Gamma_2A_{+}^2+4\kappa^2\Gamma_{+}\right), \nonumber \\
\rho_{11}&=& 4A\kappa^2, \nonumber \\
N_0&=&{A_{+}^2\Big(4\kappa^2+\Gamma_2(A+\Gamma_1)\Big)}.
\end{eqnarray}

\subsection{Liouvillian eigenmatrix $\hat\rho_{10}$ in Eq.~(\ref{r01})}

The elements of the traceless eigenmatrix $\hat\rho_{10}$, in
Eq.~(\ref{r01}), become
\begin{equation} 
\hat\rho_{10}={\rm diag}(1,-1,-1,1).
\end{equation}
$\ $

\subsection{Liouvillian eigenmatrices $\hat\rho_{3,4}$ in Eq.~(\ref{r23})} 
The elements of the traceless eigenmatrices $\hat\rho_{3,4}$,
given in Eq.~(\ref{r23}), take the form
\begin{eqnarray}  \label{C4}
\rho'_{00}\pm f_1D&=&-8\kappa^2\Gamma_{-}-\Gamma_2A_-^2\pm\Gamma_2A_-D, \nonumber \\
 \rho'_{01}\pm f_2D&=&-4\kappa^2(A-\Gamma_{+})\pm4\kappa^2D, \nonumber \\
\rho'_{0110}\pm{f_3}{D}&=& -2\kappa(\Gamma_2A_-+8\kappa^2)\pm 2\kappa\Gamma_2D, \nonumber \\
\rho'_{10}\pm f_4D&=&\Gamma_2A_-^2-4\kappa^2(A-\Gamma_1+3\Gamma_2) \nonumber \\
&&\pm(\Gamma_2^2-\Gamma_2(A+\Gamma_1)+4\kappa^2)D,\nonumber \\
\rho_{11}&=& 8A\kappa^2.
\end{eqnarray}

\subsection{Liouvillian eigenmatrix $\hat\rho_{5}$ in Eq.~(\ref{r56})} 
The elements of the traceless Hermitian eigenmatrix $\hat\rho_5$, given in Eq.~(\ref{r56}), are written as follows 
\begin{eqnarray} 
&\rho_{00}=-8\Gamma_+\kappa^2, \quad \rho_{01}=\rho_{10}=-4\kappa^2(A-\Gamma_+),& \nonumber \\
& \rho_{0110}=-\kappa\Big[4A\Gamma_1+(A-\Gamma_+)(A-\Gamma_-)\Big], \quad \rho_{11}=8A\kappa^2.& \nonumber \\
\end{eqnarray}
\subsection{Liouvillian eigenmatrices $\hat\rho_{1,2,8,9}$ in Eq.~(\ref{r4567})}
The elements of the traceless eigenmatrices $\hat\rho_{1,2,8,9}$,
given in Eq.~(\ref{r4567}), have the following forms:
\begin{eqnarray}\label{C5}  
\rho_{0001}&=&4\kappa\left[\pm E_{\pm}(\Gamma_+-A)\pm F+A_+^2-4A\Gamma_2\right], \nonumber \\
\rho_{0010}&=&\pm E_{\pm}\left(\Gamma_2^2-(A+\Gamma_1)^2\pm F\right)\pm 2\Gamma_2F \nonumber \\
&&+2\Gamma_2(\Gamma_2^2-(A+\Gamma_1)^2)+16\kappa^2(A-\Gamma_+), \nonumber \\
\rho_{0111}&=&32A\kappa^2, \nonumber \\
\rho_{1011}&=&8A\kappa(2\Gamma_2\pm E_{\pm}),
\end{eqnarray}
where $E_{\pm}$ and $F$ are given in Eq.~(\ref{lambda}).

The eigenmatrices $\hat\rho_{1,2}$ have the elements given in
Eq.~(\ref{C5}) with $E_{\pm}$ and $\pm F$, respectively. The
eigenmatrices $\hat\rho_{8,9}$ have the elements given in
Eq.~(\ref{C5}) with $-E_{\pm}$ and $\pm F$, respectively.

\section{Hermitian pseudo-eigenmatrices $\hat\rho'_{5}$ and $\hat\rho''_{5}$}
The generalized pseudo-eigenmatrices $\hat\rho'_{5}$ and $\hat\rho''_{5}$ can be found from the eigenmatrix $\hat\rho_{5}$, given in Eq.~(\ref{r56}), by applying Jordan chain relations, i.e.,
\begin{eqnarray} \label{D1}
{\cal L}\hat\rho_5-\lambda_{\rm LEP}\hat\rho_5&=&0, \nonumber \\
{\cal L}\hat\rho'_5-\lambda_{\rm LEP}\hat\rho'_5&=&\hat\rho_5, \nonumber \\
{\cal L}\hat\rho''_5-\lambda_{\rm LEP}\hat\rho''_5&=&\hat\rho'_5.
\end{eqnarray}
By combining together Eqs.~(\ref{r56}) and (\ref{D1}), one can straightforwardly arrive at the pseudo-eigenmatrices $\hat\rho'_5$ and $\hat\rho''_5$, 
which have the following general form:
\begin{equation}\label{D2}
\hat\rho^j_5\equiv\begin{pmatrix}
a^j & 0 & 0 & 0 \\
0 & b^j & \alpha^j & 0 \\
0 & \alpha^j & c^j & 0 \\
0 & 0 & 0 & d^j
\end{pmatrix}, \quad j=\{',''\}.
\end{equation}

The elements of the pseudo-eigenmatrix $\hat\rho'_5$ have the following form:
\begin{eqnarray}\label{D3}
&a'=2\Gamma_2A_--2\Gamma_+, \quad b'=\frac{1}{2}A_-^2-(A-\Gamma_+),& \nonumber \\
&c'=-\frac{1}{2}A_+^2+2\Gamma_2^2-(A-\Gamma_+), \quad d'=2A,& \nonumber \\
&\alpha'=\Gamma_2A_--A_+.&
\end{eqnarray}
And the elements of the pseudo-eigenmatrix $\hat\rho''_5$ read as follows
\begin{eqnarray}\label{D4}
&a''={\frac {6\,{{ \Gamma_2}}^{2}+ \left( -6\,A-4\,{\Gamma_1}+8 \right) 
{ \Gamma_2}-2\,{\Gamma_1}\, \left( A+{\Gamma_1} \right) }{A_-}}, &\nonumber \\
&c''=\frac{-5\,{{ \Gamma_2}}^{2}+ \left( 6\,A+4\,{ \Gamma_1}-6 \right) { 
\Gamma_2}-{A}^{2}+{{\Gamma_1}}^{2}-2\,A-2\,{ \Gamma_1}}{A_-},& \nonumber \\
&b''=2-(A-\Gamma_+), \quad d''=2A, \quad \alpha''=-\frac{-2\Gamma_2(A_--2)}{A_-}.& \nonumber \\
\end{eqnarray}
It is assumed that all elements of the pseudo-eigenmatrices $\hat\rho'_5$ and $\hat\rho''_5$, given in Eqs.~(\ref{D3}) and (\ref{D4}), respectively, have the same dimensionality.

The eigenstates of these Hermitian pseudo-eigenmatrices, which describe the intercavity interaction, become of the form:
\begin{equation}
|\psi_5^j\rangle_{\pm}\equiv2\alpha^j|10\rangle+\left(c^j-b^j\pm\sqrt{4(\alpha^j)^2+(b^j-c^j)^2}\right)|01\rangle,
\end{equation}
with $j=\{',''\}$, and where $\alpha^j$, $b^j$, and $c^j$ are given in Eqs.~(\ref{D3}) and (\ref{D4}).


%

\end{document}